\documentclass[12pt]{article}

\usepackage[margin=1in]{geometry}
\usepackage{setspace}
\usepackage{authblk} 
\usepackage{graphicx}
\graphicspath{{Fig/}}
\usepackage{subcaption}

\usepackage{amsmath, amssymb, amsthm, bm}
\usepackage{mathtools}
\usepackage{booktabs}
\usepackage{enumitem}
\usepackage{hyperref}
\usepackage{siunitx}
\usepackage{algorithm}
\usepackage[noend]{algpseudocode}
\usepackage{multirow}
\usepackage{tabularx}
\usepackage{xr}
\usepackage{booktabs}
\usepackage{makecell}
\usepackage{float}
\usepackage{tcolorbox}
\tcbuselibrary{listings}
\usepackage{xcolor}

\newcommand{\bmeta}{\boldsymbol{\beta}}
\newcommand{\bgamma}{\boldsymbol{\gamma}}

\newcommand{\cT}{\mathcal{T}}

\title{RESOLVE-IPD: High-Fidelity Individual Patient Data Reconstruction and Uncertainty-Aware Subgroup Meta-Analysis}

\author[1$\dagger$]{Lang Lang}
\author[1$\dagger$]{Yao Zhao}
\author[1]{Qiuxin Gao}
\author[1,2,*]{Yanxun Xu}

\affil[1]{\small Department of Applied Mathematics and Statistics, Johns Hopkins University}
\affil[2]{\small Division of Quantitative Sciences, Department of Oncology, Johns Hopkins University School of Medicine}
\affil[$\dagger$]{\small These authors contributed equally}
\affil[*]{\small Corresponding author: \texttt{yanxun.xu@jhu.edu}}

\date{}

\begin{document}

\maketitle

\begin{abstract}
Individual patient data (IPD) from oncology trials are essential for reliable evidence synthesis but are rarely publicly available, necessitating reconstruction from published Kaplan-Meier (KM) curves. Existing reconstruction methods suffer from digitization errors, unrealistic uniform censoring assumptions, and the inability to recover subgroup-level IPD when only aggregate statistics are available. To address these limitations, we developed RESOLVE-IPD, a unified computational framework comprising two primary components: high-fidelity IPD reconstruction and uncertainty-aware subgroup recovery. The first component integrates the VEC-KM and CEN-KM modules; VEC-KM extracts precise coordinates and explicit censoring marks from vectorized figures to minimize digitization error, while CEN-KM resolves overlapping censor symbols and eliminates the unrealistic assumption of uniform censoring. The second component utilizes the MAPLE module to identify an ensemble of data-compatible labelings that minimize discrepancies between reconstructed and reported statistics, enabling a meta-analysis framework that explicitly quantifies and propagates the uncertainty inherent in subgroup reconstruction. RESOLVE-IPD was validated through a subgroup meta-analysis of four trials in advanced esophageal squamous cell carcinoma, focusing on the programmed death ligand 1 (PD-L1)-low population. The first-stage modules reconstructed overall survival IPD that precisely matched published KM curves and reported summary statistics. MAPLE successfully inferred subgroup labels and produced plausible IPD reconstructions consistent with reported hazard ratios and survival estimates. Propagating meta-analysis confirmed a survival benefit for immunotherapy over chemotherapy, with significant effects during the 6- to 12-month period.
\end{abstract}

\section*{Introduction}
Individual patient data (IPD) from oncology trials are essential for a wide range of secondary analyses for evidence synthesis, including the identification of prognostic biomarkers, assessment of treatment effect heterogeneity, and development of clinical practice guidelines \cite{debray2015individual,smith2016individual,riley2010meta}.  These analyses critically depend on time-to-event endpoints like overall survival (OS) and progression-free survival (PFS). Direct IPD access is essential not only to perform these analyses with maximum flexibility but also to validate core modeling assumptions, such as proportional hazards, which are frequently challenged in trials of immunotherapies and targeted agents \cite{batson2016review}. For these reasons, IPD meta-analysis is regarded as the gold standard for evidence synthesis and has a direct impact on clinical decision-making. Despite its importance, access to proprietary IPD is often restricted due to confidentiality, governance, and resource constraints. Consequently, secondary analyses typically rely on either synthesizing published aggregate summary statistics or reconstructing IPD from publicly available Kaplan-Meier (KM) curves.

The accuracy and utility of such secondary analyses are fundamentally constrained by limitations in existing IPD reconstruction methods \cite{guyot2012,Liu2021,zhao2022kmsubtraction,rogula2022method,zhao2025synthipd}. These limitations stem from two main sources: errors in data acquisition and structural flaws in the underlying statistical algorithms.
First, traditional approaches rely on manual digitization or low-resolution raster images, introducing coordinate extraction errors. For instance, tools like IPDfromKM \cite{Liu2021} require manual calibration, which is both labor-intensive and error-prone. Although newer methods like KM-GPT \cite{ZhaoSunDingXu2025} leverage computer vision for automated coordinate detection, residual inaccuracies remain. Second, the dominant reconstruction algorithm, namely the modified iterative Kaplan-Meier (iKM) method, assumes a uniform censoring mechanism within intervals, as it does not account for censoring marks during reconstruction~\cite{Liu2021}. This assumption is clinically unrealistic, as censoring usually occurs at scheduled assessments or due to patient dropout, leading to biased survival estimates. The KMtoIPD algorithm \cite{rogula2022method} introduced an improvement by allowing digitized censoring times to be incorporated directly into the reconstruction, but it overlooks the issue of overlapping censor marks, in which a single symbol can correspond to multiple censored patients. This limitation results in underestimation of censoring frequency and degraded reconstruction fidelity, particularly under high censoring.  More recently, SynthIPD \cite{zhao2025synthipd} digitizes KM curves at vector-level precision and extracts event and censoring coordinates. Nevertheless, its IPD reconstruction step still estimates the number of events within each risk-table–aligned interval using an allocation approach similar to that of the modified iKM algorithm. As a result, when the risk table is sparse or the censoring rate is high, the interval-specific event estimates may be less reliable, which can limit the accuracy of the reconstructed IPD. These inaccuracies propagate through all subsequent analyses, systematically biasing the estimation of key clinical endpoints such as the hazard ratio and restricted mean survival time (RMST). 

These flaws are not just theoretical; they have consequences in real-world secondary analyses, where the reliability of derived evidence directly impacts clinical decision-making. For example, a secondary analysis~\cite{yu2022association} that relied on reconstructed IPD to assess biomarker-defined subgroups (e.g., the programmed death ligand 1 (PD-L1) low expression population in KEYNOTE-048~\cite{burtness2019pembrolizumab}) produced survival results that were substantially inconsistent with other published data for the same subgroup~\cite{schoenfeld2020keynote}. Such discrepancies underscore how even modest reconstruction errors can bias secondary analyses, potentially misleading clinical interpretations and evidence-based recommendations.

A further critical challenge arises in the era of precision oncology, where subgroup-based (e.g., older versus younger adults) or biomarker-stratified (e.g., PD-L1 expression) meta-analyses are essential.  While trial publications typically provide KM curves for the overall study population, results for critical subgroups are often reported only as summary statistics: a median overall survival (mOS) and a hazard ratio (HR) with their confidence intervals (CI). The absence of subgroup-specific survival curves creates a major analytical bottleneck. The core difficulty lies in non-identifiability: reconstructing individual subgroup labels for hundreds of patients from a few aggregate metrics is an underdetermined problem with a vast number of valid solutions. Numerous patient-level assignments can reproduce the reported mOS and HR equally well, yielding no unique reconstruction. This intrinsic ambiguity constitutes a fundamental source of epistemic uncertainty. For any subsequent meta-analysis to be reliable, this uncertainty must be explicitly quantified and transparently propagated throughout the evidence synthesis process.

To address these methodological and clinical challenges,  we develop {\bf RESOLVE-IPD}: a unified computational framework for high-fidelity IPD reconstruction and robust downstream evidence synthesis, with a particular focus on uncertainty-aware subgroup meta-analyses. RESOLVE-IPD addresses critical limitations in two key areas. First, it resolves the accuracy bottleneck by combining high-fidelity vector graphics extraction with a novel IPD reconstruction algorithm designed to utilize explicit censoring marks, thereby eliminating the systemic bias of the uniform censoring assumption. Second, to recover unreported subgroup information, we introduce an optimization-based method that leverages published summary statistics to infer missing subgroup survival curves from the overall reconstructed IPD. This methodology identifies the set of data-compatible subgroup label assignments, defined as those consistent with the published summary statistics, and develops rigorous techniques to quantify, report, and systematically propagate the resulting uncertainty into downstream meta-analyses. Collectively, RESOLVE-IPD establishes a new standard for secondary data analysis in oncology by ensuring statistically robust IPD recovery and enabling novel, biomarker-stratified subgroup analyses with fully quantified uncertainty.

\section*{Methods}

\noindent\textbf{\textit{Overview of the RESOLVE-IPD Framework}}

\begin{figure}[htbp!]
    \centering
    \includegraphics[width=0.95\linewidth]{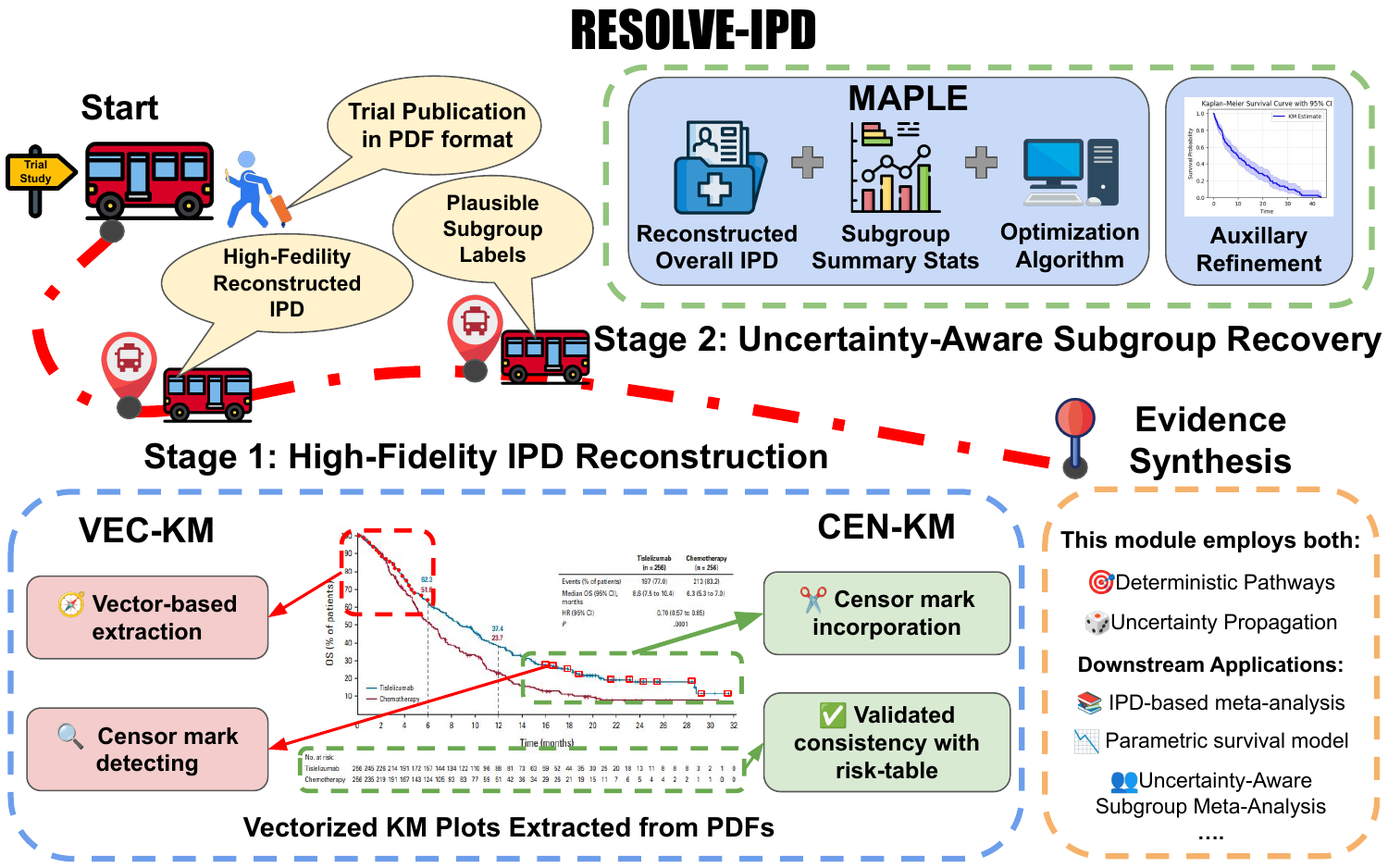}
    \caption{Schematic overview of the RESOLVE-IPD framework.}
    \label{fig_pipeline_overview}
\end{figure}

RESOLVE-IPD is a unified computational workflow designed to overcome the major limitations in reconstructing IPD from oncology trial publications. Its architecture, illustrated in Fig~\ref{fig_pipeline_overview}, consists of two synergistic stages to ensure the reliability of downstream evidence synthesis.
The first stage focuses on high-fidelity IPD reconstruction. It begins with the VEC-KM (Vector Kaplan–Meier Curve Extraction) module, which extracts precise coordinates and censoring information from KM curve images using vector graphics processing, thereby minimizing digitization error. The high-fidelity data are then processed by the CEN-KM (Censoring-Informed Kaplan–Meier) algorithm, a novel reconstruction method that correctly interprets explicit and overlapping censoring marks. This approach eliminates the systematic bias introduced by the conventional assumption of uniform censoring within intervals.
The second stage addresses uncertainty-aware subgroup recovery. When trial publications do not include KM curves for clinically relevant subgroups, such as those defined by biomarkers like PD-L1, RESOLVE-IPD applies the MAPLE (Marginal Assignment of Plausible Subgroup Labels and Evidence Propagation) algorithm. This module infers plausible subgroup labels for the reconstructed IPD using available summary statistics, such as mOS and HRs. Recognizing the inherent non-identifiability of this inverse problem, MAPLE does not seek a single definitive solution but instead characterizes the full space of label configurations consistent with the published evidence, thereby quantifying the uncertainty in subgroup-level survival estimates.
The outputs from these two stages feed directly into a downstream evidence synthesis module, which rigorously propagates the quantified uncertainty into the final meta-analytic estimates, such as pooled HR or RMST. This end-to-end design ensures that conclusions drawn from secondary analyses are statistically robust, transparent, and clinically reliable. The following sections detail each core component.

\noindent\textbf{\textit{High-Fidelity IPD Reconstruction}}

This stage integrates two components: VEC-KM for digitizing published KM plots and CEN-KM for reconstructing IPD from the digitized data. Together, they provide higher fidelity digitization and reconstruction than existing approaches.

VEC-KM addresses digitization errors by extracting vector graphics from published KM figures, using scalable vector graphics (SVG) files from journal websites or native vector paths embedded in PDF articles. As many medical journals now require figure submission in vector format, these files preserve precise geometric information. Traditional digitization methods that operate on raster images inevitably introduce rounding and alignment errors, depend on image resolution, and often require manual calibration. In contrast, vector graphics encode KM curves and censoring marks as mathematical coordinates and drawing instructions, enabling resolution-independent extraction of coordinates that accurately reflect the original plotting geometry. By parsing these vector coordinates directly, VEC-KM yields event times and survival probabilities with high numerical precision (up to six decimal places in our empirical evaluation), substantially improving accuracy over conventional pixel-based or manual approaches.

The core functionality of VEC-KM is the precise identification of both event coordinates (the stepping points of the KM curve) and censoring marks. This process is executed in three sequential steps. First, KM curve extraction identifies all line segments within the figure and selects the top $K$ longest connected paths, where $K$ is the number of KM curves, as the target survival functions. Second, axis calibration locates the $x$- and $y$-axes by locating horizontal and vertical lines whose lengths closely match those of the extracted KM curves and that lie nearest to them in perpendicular distance.  
Third, censor mark detection isolates the remaining short vector segments, classifying them as censoring marks based on their geometric shape and proximity to the corresponding KM curve. Each mark is then orthogonally projected onto the time axis to determine its exact censoring time.  This direct extraction of precise censoring times is the critical advancement that eliminates the reliance on the statistically flawed uniform censoring assumption in subsequent IPD reconstruction.

The CEN-KM module utilizes the high-fidelity outputs of VEC-KM to reconstruct the IPD, directly addressing the fundamental statistical weakness of existing algorithms. First, instead of assuming that censoring occurs uniformly within each time interval, it uses the explicit censoring times identified by VEC-KM, preserving the true censoring pattern reported in the trials. Second, it adjusts the number of censors at each time point to accommodate overlapping censor marks, a common scenario in clinical trial KM plots where a single censor symbol may represent multiple censored patients. Together, these improvements enable a more accurate reconstruction of both the event and censoring data, thereby improving the precision of survival curve recovery and subsequent analyses.

The algorithm begins with the set of event intervals derived from the KM curve and the explicit censoring times. Within each interval, the algorithm first assigns a minimum number of censor events by counting the unique censor marks present. The algorithm then identifies candidate event counts that close to the theoretical value derived from the KM equation, testing the integers which possibly reproduce the observed survival drop. For each candidate event count, the model iteratively increments the number of overlapping censors until the difference between the estimated and target survival probability falls below a predefined precision threshold. The event-censor combination that yields the smallest deviation is selected as the final reconstruction for that interval. When the algorithm reaches a time point tied to the number-at-risk reported in the risk table, it performs a targeted alignment step to ensure that the reconstructed population at risk exactly matches the published value. To achieve this, the algorithm adjusts the counts of censors to correct discrepancies while preserving the survival structure.

The branch-wise comparison of candidates with iterative censor adjustment simultaneously resolves the issue of overlapping censor marks and corrects for integer rounding errors that commonly arise in event number estimation. By systematically evaluating multiple plausible event-censor configurations rather than relying on a single rounded estimate, CEN-KM ensures the reconstructed survival probability faithfully mirrors the published KM curve. Combining precise censoring data, dynamic event-censor adjustments, and risk-table alignment, the CEN-KM algorithm reconstructs time-to-event data with exceptional accuracy, establishing a robust foundation for downstream survival analyses and meta-analyses. Comprehensive details about the VEC-KM and CEN-KM algorithms can be found in Supplementary Material Section~\ref{sec_HF_IPD_detail}.

\noindent\textbf{\textit{Uncertainty-Aware Subgroup IPD Recovery}}

Clinical trial publications vary considerably in how they report subgroup data. Some provide KM curves for key subgroups, while others report only aggregate statistics such as mOS, HRs, and sample sizes. The RESOLVE-IPD framework is designed to account for this variability, offering two complementary pathways to recover subgroup-level IPD, depending on the evidence available. When subgroup KM curves are provided, a deterministic reconstruction pathway assigns labels directly, ensuring the highest precision. Conversely, when only summary statistics are available, RESOLVE-IPD employs its optimization-based MAPLE algorithm (Marginal Assignment of Plausible Subgroup Labels and Evidence Propagation) to infer subgroup labels consistent with the published constraints. Since recovering exact subgroup labels solely from summary statistics is fundamentally non-identifiable, RESOLVE-IPD quantifies the uncertainty by an ensemble of plausible labelings, which can be propagated into downstream analyses to support robust inference.

The deterministic pathway is used when high-fidelity KM figures, available in vector format, contain sufficient detail for direct reconstruction. This occurs in two scenarios: (i) when the KM curve for the target subgroup is explicitly reported, or (ii) when high-fidelity overall and subgroup KM curves are available for all but one subgroup. In the first scenario, the VEC-KM and CEN-KM modules are applied directly to the target subgroup curve for precise reconstruction.  In the second, the missing subgroup's IPD is derived using a KM subtraction procedure applied to the reconstructed high-fidelity IPD obtained from the available overall and subgroup KM plots~\cite{zhao2022kmsubtraction}. A detailed description of the KM subtraction procedure is provided in the Supplementary Material Section~\ref{sec_kmsub_detail}. This deterministic pathway yields the most reliable subgroup IPD and serves as the benchmark for accuracy. 

When high-fidelity graphical data are unavailable, RESOLVE-IPD employs its optimization-based MAPLE pathway.  MAPLE begins by reconstructing the overall IPD and assigns each patient $i$ a binary or categorical subgroup label~$g_i$ (e.g., high vs.\ low biomarker expression) by solving an optimization problem. The objective is to find labelings such that the reconstructed stratified IPD reproduces the reported summary statistics $\boldsymbol{\tau}^{\mathrm{tgt}}$, which typically include HR, mOS, and their CIs. 

Since this inverse problem is inherently non-identifiable, summary statistics often admit many distinct yet data-compatible subgroup labelings. To address this, MAPLE first finds configurations with the smallest recovery error, denoted~$\mathcal{G}_{\mathrm{MAPLE}}$. 
When publications include subgroup KM plots only as low-fidelity raster images (e.g., in supplementary materials), the algorithm leverages existing tools like KM-GPT~\cite{ZhaoSunDingXu2025} to reconstruct approximate IPD. MAPLE then uses this low-fidelity data to refine the candidate set, filtering labelings to retain only those whose derived subgroup KM curves fall within the 95\% CI of the reconstructed curve, denoted~$\mathcal{G}_{\mathrm{MAPLE}}^*$. This ensures alignment with all available evidence while acknowledging and preserving the inherent uncertainty stemming from non-identifiability.

The output of MAPLE is an ensemble of plausible subgroup labelings, denoted~$\mathcal{G}_{\mathrm{MAPLE}}$ or $\mathcal{G}_{\mathrm{MAPLE}}^*$ when further refined. Each member of the ensemble represents a distinct yet data-compatible reconstruction of subgroup-level IPD, offering a principled quantification of uncertainty in subgroup assignment. This ensemble provides a principled quantification of uncertainty in subgroup assignment and forms the basis for uncertainty-aware downstream analyses. Detailed optimization formulations and algorithmic implementations of MAPLE are outlined in Supplemental Material Section~\ref{sec_maple_detail}. 

\noindent\textbf{\textit{Robust Meta-Analysis with Propagated Uncertainty}}

The ensemble of subgroup labels generated by MAPLE unlocks a broad range of downstream analyses, such as subgroup-specific treatment effect estimation and meta-analysis stratified by clinically meaningful subgroups. In this paper, we focus on subgroup meta-analysis and demonstrate how RESOLVE-IPD propagates the uncertainty from non-identifiable subgroup labels into the final evidence synthesis. 

Consider a meta-analysis of \(K\) studies. For each study \(k\), the RESOLVE-IPD pipeline produces an ensemble of plausible subgroup labelings, denoted \(\mathcal{G}_{\mathrm{MAPLE}}^{*(k)}\). Conventional meta-analyses typically use a single, fixed dataset per study, thereby ignoring the uncertainty introduced during IPD reconstruction. In contrast, our framework performs a propagating meta-analysis that explicitly accounts for this uncertainty. Specifically, the meta-analysis is repeatedly executed on randomly sampled subgroup datasets, with one labeling selected from each study's ensemble \(\mathcal{G}_{\mathrm{MAPLE}}^{*(1)}, \ldots, \mathcal{G}_{\mathrm{MAPLE}}^{*(K)}\). Each iteration produces a set of pooled estimates, such as the combined HR and its CI. Repeating this process multiple times yields a Monte Carlo distribution of the pooled estimates. The outcome is no longer necessarily a single pooled HR but a distribution over the summary measures of interest. Depending on the analysis, these may include the mOS, pooled HR, or time-varying HR$(t)$ when proportional hazards do not hold. We propose reporting summary measures of all estimates and their associated CIs resulting from each Monte Carlo run.

This approach provides a transparent and quantitative measure of evidence robustness. A narrow distribution indicates that the meta-analytic conclusions are stable despite label uncertainty, whereas a wide distribution reveals sensitivity to subgroup ambiguity, an important consideration for evidence grading and clinical guideline development. By rigorously quantifying and propagating these uncertainties, RESOLVE-IPD strengthens the statistical validity and credibility of meta-analytic evidence in oncology. The implementation code for RESOLVE-IPD, including VEC-KM, CEN-KM, and MAPLE modules, applied to simulation data, is publicly available at: \url{https://github.com/JackZhao0312/RESOLVE-IPD}. The CEN-KM module is also available via a web-based interface at \url{https://km-gpt.wse.jhu.edu/}.

\section*{Results}

\noindent\textbf{\textit{Evaluation of RESOLVE-IPD on Synthetic Data}}

We first assessed the performance of RESOLVE-IPD using a simulated dataset with stratified survival outcomes (biomarker-high vs. biomarker-low) to evaluate the pipeline's reconstruction capabilities. The simulation setup is detailed in Supplementary Material Section~\ref{sec_sim_generation}. We evaluated both the high-fidelity IPD reconstruction stage (VEC-KM + CEN-KM) and the subgroup label recovery stage.

\begin{figure}[htbp!]
    \centering
    \includegraphics[width=\linewidth]{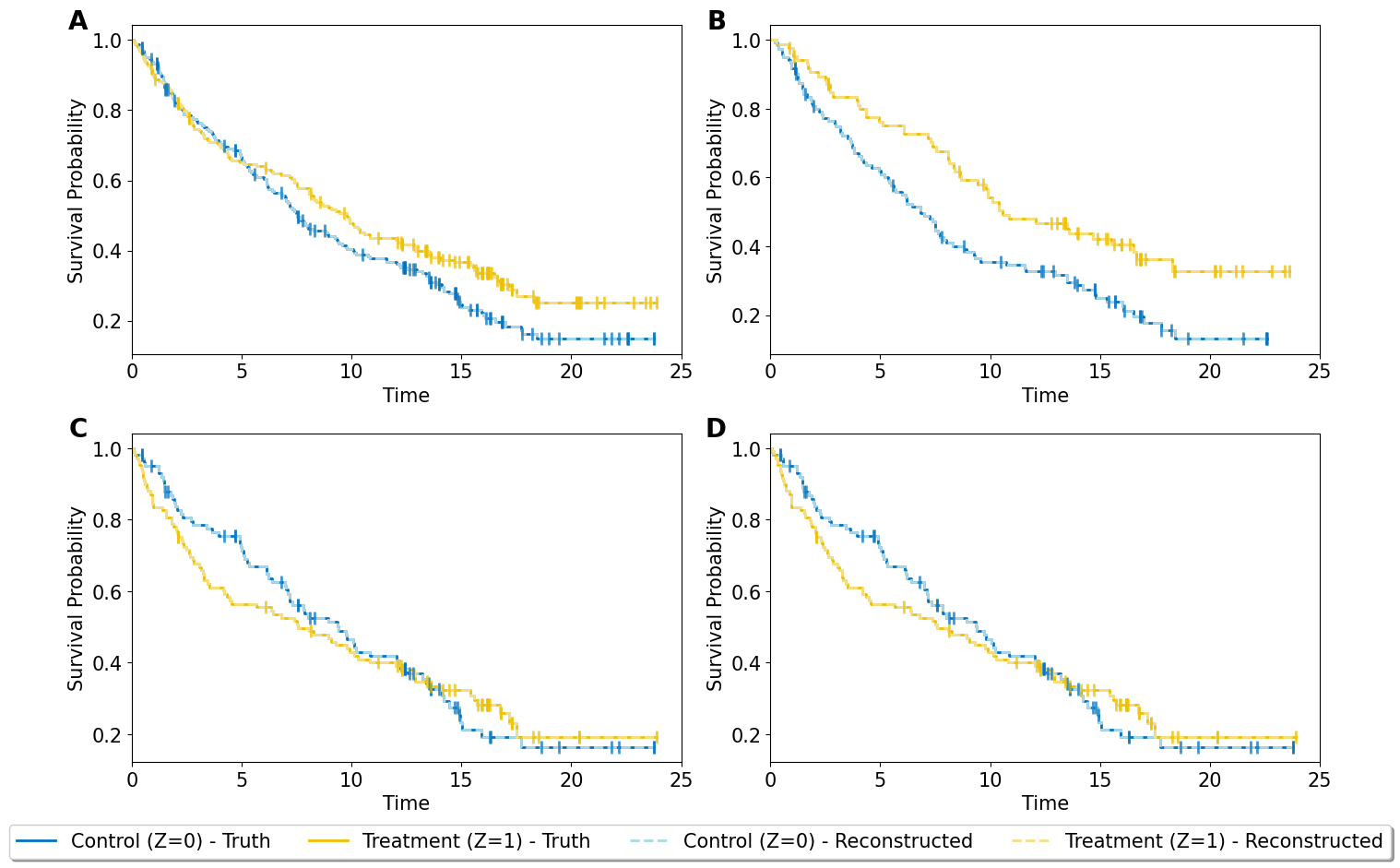}
    \caption{KM plots derived from reconstructed subgroup-level IPD using the VEC-KM + CEN-KM pipeline, overlaid with the ground-truth KM curves. (A) Overall population. (B) Biomarker-high subgroup. (C) Biomarker-low subgroup reconstructed directly using VEC-KM + CEN-KM. (D) Biomarker-low subgroup reconstructed via the KM subtraction method. }
    \label{fig_sim_cenkm}
\end{figure}

To benchmark reconstruction accuracy, we reconstructed IPD from simulated KM plots for the overall trial population and all subgroups, comparing the reconstructed results against the true simulated IPD. The VEC-KM + CEN-KM pipeline achieved perfect fidelity, recovering the ground-truth IPD with a root mean square error (RMSE) of 0 for survival probabilities across all subgroups and treatment arms. To demonstrate CEN-KM's advantages over existing methods that assume uniform censoring or fail to account for multiple censored patients at single censoring marks, we compared its performance against IPDfromKM and KMtoIPD using perfectly digitized KM data from VEC-KM as input. Both alternative methods demonstrated small but consistent errors (RMSE range: 0.002–0.009, Supplementary Table~\ref{tab_sim_reconstruction_error}), reflecting limitations in their handling of censoring. By contrast, the results from VEC-KM + CEN-KM achieved perfect alignment with the ground truth (Fig~\ref{fig_sim_cenkm}A-C), while visual comparisons revealed noticeable inaccuracies for IPDfromKM and KMtoIPD (Supplementary Fig~\ref{fig_sim_ipdfromkm}A-C and~\ref{fig_sim_kmtoipd}A-C). These results validate VEC-KM's exceptional precision in curve digitization and highlight CEN-KM's capability to accurately reconstruct IPD by leveraging explicit censoring information.

To evaluate subgroup label recovery, we initially assessed the performance of the deterministic KM subtraction method in scenarios where high-fidelity KM curves were available for the overall cohort and all but one subgroup. Specifically, we withheld the KM curve for the low biomarker expression subgroup. The KM subtraction method successfully achieved exact recovery of the ground truth (RMSE = 0, Fig~\ref{fig_sim_cenkm}D). In contrast, benchmark tools demonstrated observable deviations from the ground truth (Supplementary Fig~\ref{fig_sim_ipdfromkm}D and~\ref{fig_sim_kmtoipd}D).

\begin{figure}[htbp!]
    \centering
    \includegraphics[width=0.85\textwidth]{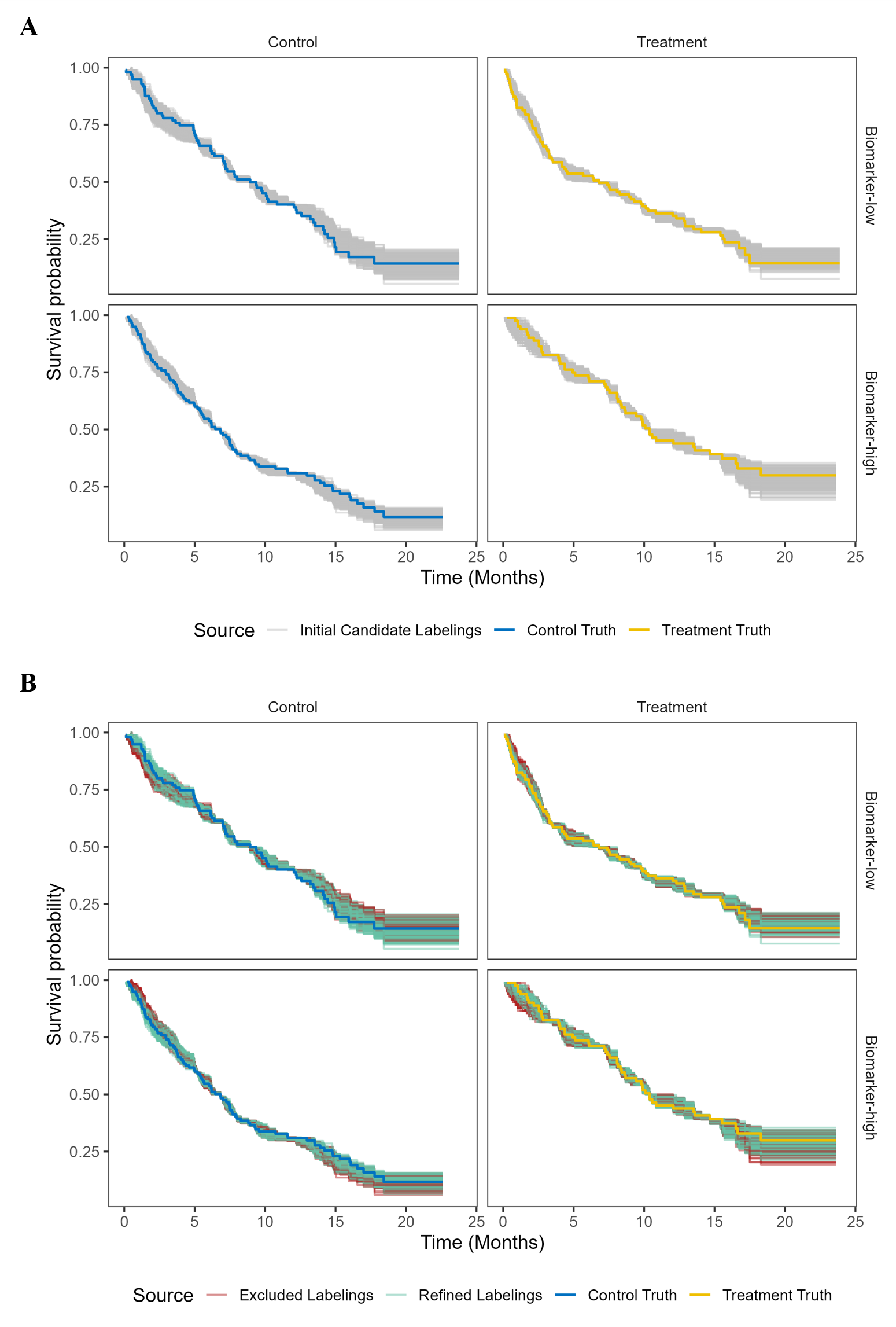}
    \caption{True subgroup KM curves overlaid with (A) reconstructed KM curves from all initial candidate labelings $\mathcal{G}_{\mathrm{MAPLE}}$, and (B) the refined set of plausible labelings  $\mathcal{G}_{\mathrm{MAPLE}}^*$, showing excluded (red) and retained (green) labelings.}
    \label{fig_sim_maple}
\end{figure}

Next, we evaluated the MAPLE module's ability to infer patient-level subgroup labels ($g_i$) using reconstructed high-fidelity IPD from the overall population. The optimization algorithm was applied based on aggregate summary statistics, including subgroup-level HRs and mOS with CIs, along with counts of patients by treatment arm, subgroup, and event status derived from synthetic data. Across 1,000 optimization runs, MAPLE identified 809 unique subgroup labelings, denoted  $\mathcal{G}_{\mathrm{MAPLE}}$, each exactly reproducing the target subgroup-level summary statistics. This large ensemble of mathematically equivalent solutions demonstrates the inherent high-dimensional non-identifiability of the subgroup assignment problem. Fig~\ref{fig_sim_maple}A presents the true KM curves (blue lines for control group and yellow lines for treatment group) overlaid with KM curves derived from all initial candidate labelings in $\mathcal{G}_{\mathrm{MAPLE}}$ (grey lines).  The true subgroup KM curves lie within the gray-shaded reconstructed region, demonstrating that MAPLE effectively accounts for uncertainty while ensuring the true survival patterns are preserved within the reconstructed ensemble.

Finally, we evaluated MAPLE's ability to refine the labeling ensemble by leveraging low-fidelity raster images of subgroup KM curves, as commonly presented in publications. Labelings were filtered to retain only those whose high-subgroup KM curves fell within the true high-subgroup KM curve's 95\% CI, resulting in a refined set of 603 labelings, denoted as $\mathcal{G}_{\mathrm{MAPLE}}^*$. This refinement excluded unreliable labelings while preserving solutions closely aligned with the true survival patterns of the low biomarker expression subgroup (Fig~\ref{fig_sim_maple}B).

\noindent\textbf{\textit{Application to Real-World Oncology Data}}

The high-fidelity, uncertainty-aware IPD reconstruction provided by the RESOLVE-IPD pipeline establishes a new foundation for robust secondary analysis. To demonstrate its clinical utility, we considered a subgroup meta-analysis evaluating the efficacy of immunotherapy versus chemotherapy as a second-line treatment for advanced esophageal squamous cell carcinoma (ESCC), with a particular focus on the PD-L1 low-expression subgroup.  We analyzed data from four pivotal clinical trials: RATIONALE-302~\cite{shen2022tislelizumab}, KEYNOTE-181~\cite{doi2016keynote}, ESCORT~\cite{huang2020camrelizumab}, and ATTRACTION-3~\cite{kato2019nivolumab}, which were identified through a systematic literature search and assessed for risk of bias~\cite{yap2023}. The RESOLVE-IPD framework was applied to all trials to reconstruct IPD and recover subgroup labels for subsequent meta-analysis. 

SVG-format KM plots for the overall population were available for all four trials. We validated the reconstruction pipeline using OS data from these trials.  The IPD reconstructed by the VEC-KM and CEN-KM modules exactly reproduced the published KM curves, with overlaid plots showing perfect alignment (Supplementary Fig~\ref{fig_meta_keynote_overall_overlay} -~\ref{fig_meta_rationale_overall_overlay}). All derived summary statistics, including mOS and HRs, were identical to the originally reported values (Supplementary Table~\ref{tab_meta_overall_assess}), establishing a reliable foundation for subsequent analyses. 

\begin{figure}[htbp!]
    \centering
    \includegraphics[width=\textwidth]{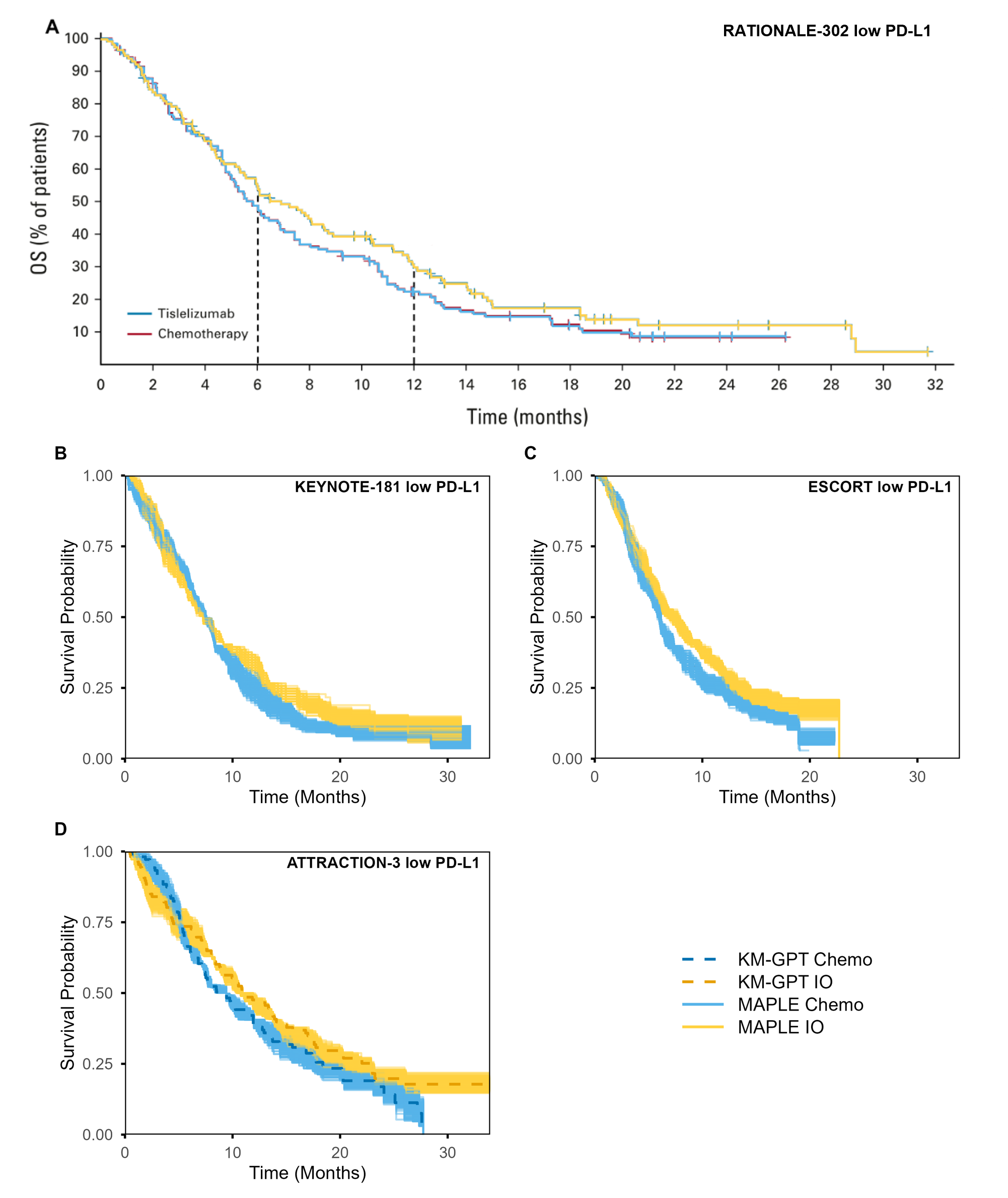}
    \captionsetup{font=small}
    \caption{Reconstructed KM curves overlaid on the original publication curve for RATIONALE-302 and on KM-GPT–derived curves extracted from low-fidelity raster images for the remaining studies. (A) RATIONALE-302 PD-L1–low subgroup reconstructed via the High-Fidelity IPD Reconstruction procedure. (B)-(D) KEYNOTE-181, ESCORT and ATTRACTION-3 PD-L1–low subgroups reconstructed by MAPLE. IO = Immunotherapy, Chemo = Chemotherapy.}
    \label{fig_meta_maple_overlay}
\end{figure}

The availability of data for the target PD-L1 low subgroup, however, was heterogeneous across trials. Among them, only RATIONALE-302 provided high-fidelity SVG-format KM curves for its PD-L1 subgroups, allowing direct IPD reconstruction. For this trial, the reconstructed KM curve for the PD-L1 low subgroup was identical to the published curve, with perfectly matched summary statistics, serving as a high-confidence benchmark (Fig~\ref{fig_meta_maple_overlay}A).

For the remaining three trials (KEYNOTE-181, ESCORT and ATTRACTION-3), we employed the MAPLE algorithm to recover patient-level subgroup assignments. The available data varied: ATTRACTION-3 had low-fidelity raster images for both PD-L1 high and low subgroups, while KEYNOTE-181 and ESCORT had only low-fidelity images for the PD-L1 high subgroup. The initial optimization process, with details on the information extracted from the original publication provided in Supplementary Table~\ref{tab_meta_maple_info}, generated a large solution space, producing 2,304, 6,736, and 7,982 candidate labelings for KEYNOTE-181, ESCORT, and ATTRACTION-3, respectively.  Each labeling represents a complete, self-consistent set of patient subgroup assignments that closely reproduces the published summary statistics. From this initial pool, we selected the subset with the smallest error. For KEYNOTE-181, all summary targets were recovered exactly; for ESCORT and ATTRACTION-3, the maximum deviation across all summary targets was approximately 1\%.

We further refined these candidate sets using the available low-fidelity graphical data. For KEYNOTE-181 and ESCORT, we used KM-GPT to reconstruct approximate KM curves from the PD-L1 high subgroup images and filtered the candidate labelings, retaining only those whose derived PD-L1 high survival curves fell within the 95\% CI of the KM-GPT reconstruction. For ATTRACTION-3, this filtering was applied simultaneously to both subgroups using the available low-fidelity images for both groups. This refinement yielded final ensembles of 425, 489, and 1,063 plausible labelings for KEYNOTE-181, ESCORT, and ATTRACTION-3, respectively. 

The summary statistics derived from the IPD based on these refined labelings, including mOS, HR, and their CIs, are presented alongside the originally published values in Table~\ref{tab_report_maple_comparison}. The reconstructed values closely align with the originally published statistics, reflecting the accuracy of RESOLVE-IPD in recovering subgroup-specific information from incomplete data. Notably, the distributions of reconstructed values within the refined labeling ensembles highlight varying degrees of uncertainty, as evidenced by the frequency of plausible subgroup estimates for trials such as ESCORT and ATTRACTION-3. These findings suggest that while refinement enhances concordance, ensemble-based variability remains a key consideration in interpreting subgroup-specific outcomes. The ensembles of KM curves for the PD-L1 low subgroups in KEYNOTE-181 and ESCORT further highlight this uncertainty (Fig~\ref{fig_meta_maple_overlay}B, C), while the ensemble for ATTRACTION-3 shows close agreement with the KM-GPT-derived curve from the low-fidelity image (Fig~\ref{fig_meta_maple_overlay}D). Similarly, the reconstructed KM curves for PD-L1 high subgroups across these trials display comparable patterns (Supplementary Fig~\ref{fig_meta_overlay_km_high}).

\begin{table}[!htbp]
\centering
\caption{Validation of reconstructed subgroup statistics against originally reported values.
Comparison of published summary statistics for PD-L1 high and low subgroups, including mOS, HRs, and their CIs, with the values reconstructed by RESOLVE-IPD. The table presents the originally reported values from the trial alongside distributions of reconstructed values across the refined labeling ensemble. Each reconstructed statistic is shown with its associated CI (in parentheses) and the proportion of labelings within the ensemble reporting that statistic (indicated as a percentage). For ESCORT, mOS information was unavailable for the PD-L1–low subgroup, precluding direct comparison.  IO, immunotherapy; Chemo, chemotherapy.}
\renewcommand{\arraystretch}{1.15}
\resizebox{\linewidth}{!}{%
\begin{tabular}{>{\centering\arraybackslash}m{0.8cm} l l c c c}
\toprule
\textbf{{Trial}} & \textbf{PD-L1} & \textbf{Treatment} & \textbf{Statistic} & \textbf{Reported} & \textbf{RESOLVE-IPD} \\
\midrule
\multirow{6}{*}{\rotatebox{90}{\textbf{KEYNOTE-181}}}
  & \multirow{3}{*}{High}
     & Chemo & mOS & 6.7 (4.8, 8.6) & 6.7 (4.8, 8.6) (100.0\%) \\
  &  & IO    & mOS & 10.3 (7.0, 13.5) & 10.3 (7.0, 13.5) (100.0\%) \\
  &  & IO vs.\ Chemo & HR & 0.64 (0.46, 0.90) & 0.64 (0.46, 0.90) (100.0\%) \\
  \cmidrule(l){2-6}
  & \multirow{3}{*}{Low}
     & Chemo & mOS & 7.5 (6.3, 8.3) & 7.5 (6.3, 8.3) (100.0\%) \\
  &  & IO    & mOS & 7.3 (5.7, 9.2) & 7.3 (5.7, 9.2) (100.0\%) \\
  &  & IO vs.\ Chemo & HR & 0.88 (0.66, 1.16) & 0.88 (0.66, 1.16) (100.0\%) \\
\midrule
\multirow{8}{*}{\rotatebox{90}{\textbf{ESCORT}}}
  & \multirow{4}{*}{High}
     & Chemo & mOS & 6.3 (5.5, 7.5) & 6.3 (5.5, 7.5) (100.0\%) \\
  &  & IO    & mOS & 9.2 (7.0, 11.2) & 9.1 (7.0, 11.1) (42.5\%), 9.1 (7.0, 11.3) (29.9\%),\\
  &  &       &     &                   & 9.1 (7.0, 11.2) (27.6\%) \\
  &  & IO vs.\ Chemo & HR & 0.58 (0.42, 0.81) & 0.58 (0.42, 0.81) (100.0\%) \\
  \cmidrule(l){2-6}
  & \multirow{3}{*}{Low}
     & Chemo & mOS & -- & -- \\
  &  & IO    & mOS & -- & -- \\
  &  & IO vs.\ Chemo & HR & 0.82 (0.62, 1.09) & 0.82 (0.62, 1.08) (82.8\%), 0.82 (0.62, 1.09) (17.2\%) \\
\midrule
\multirow{10}{*}{\rotatebox{90}{\textbf{ATTRACTION-3}}}
  & \multirow{6}{*}{High}
     & Chemo & mOS & 8.1 (6.0, 9.9) & 8.1 (6.0, 9.8) (54.0\%), 8.1 (6.0, 9.9) (46.0\%) \\
  &  & IO    & mOS & 10.9 (8.0, 14.2) & 10.9 (8.0, 14.2) (38.5\%), 10.9 (8.0, 14.1) (20.4\%),\\
  &  &       &     &                   & 10.8 (8.0, 14.2) (15.1\%), 10.8 (8.0, 14.1) (10.5\%),\\
  &  &       &     &                   & 10.9 (8.0, 14.3) (10.4\%), 10.8 (8.0, 14.3) (5.0\%) \\
  &  & IO vs.\ Chemo & HR & 0.69 (0.51, 0.94) & 0.69 (0.51, 0.94) (93.9\%), 0.69 (0.51, 0.93) (4.4\%),\\
  &  &       &     &                   & 0.69 (0.51, 0.95) (1.7\%) \\
  \cmidrule(l){2-6}
  & \multirow{5}{*}{Low}
     & Chemo & mOS & 9.3 (7.2, 12.0) & 9.4 (7.2, 12.0) (63.5\%), 9.4 (7.2, 11.9) (36.5\%) \\
  &  & IO    & mOS & 10.9 (8.4, 13.9) & 10.9 (8.4, 13.9) (45.7\%), 10.8 (8.4, 13.9) (21.7\%),\\
  &  &       &     &                   & 10.9 (8.4, 13.8) (21.4\%), 10.8 (8.4, 13.8) (11.2\%) \\
  &  & IO vs.\ Chemo & HR & 0.84 (0.62, 1.14) & 0.84 (0.62, 1.14) (74.8\%), 0.84 (0.62, 1.13) (15.0\%),\\
  &  &       &     &                   & 0.84 (0.62, 1.15) (10.3\%) \\
\bottomrule
\end{tabular}
}
\label{tab_report_maple_comparison}
\end{table}


\noindent\textbf{\textit{Subgroup Meta-Analysis}}

With the reconstructed IPD from RATIONALE-302 and the ensemble of candidate subgroup labelings $\mathcal{G}^*_{\mathrm{MAPLE}}$ of KEYNOTE-181, ESCORT and ATTRACTION-3, we conducted the evidence synthesis using a Bayesian piecewise exponential model with random effects to account for between-study heterogeneity, setting the time interval length to 6 months (methodological details are provided in Supplementary Section \ref{sec_meta_model}). To address the uncertainty in subgroup label assignments for three of the four trials, we implemented a propagating meta-analysis based on 500 Monte Carlo simulations. In each simulation, a single plausible subgroup labeling was randomly sampled from $\mathcal{G}^*_{\mathrm{MAPLE}}$ for each relevant trial to construct a complete analytical dataset. Each simulation produced 5,000 posterior samples following a 2,000-iteration burn-in phase, enabling us to estimate time-varying HRs and mOS with 95\% CIs. Results from all simulations were aggregated to fully characterize the uncertainties associated with subgroup label non-identifiability.

\begin{figure}[htbp!]
    \centering
    \includegraphics[width=\textwidth]{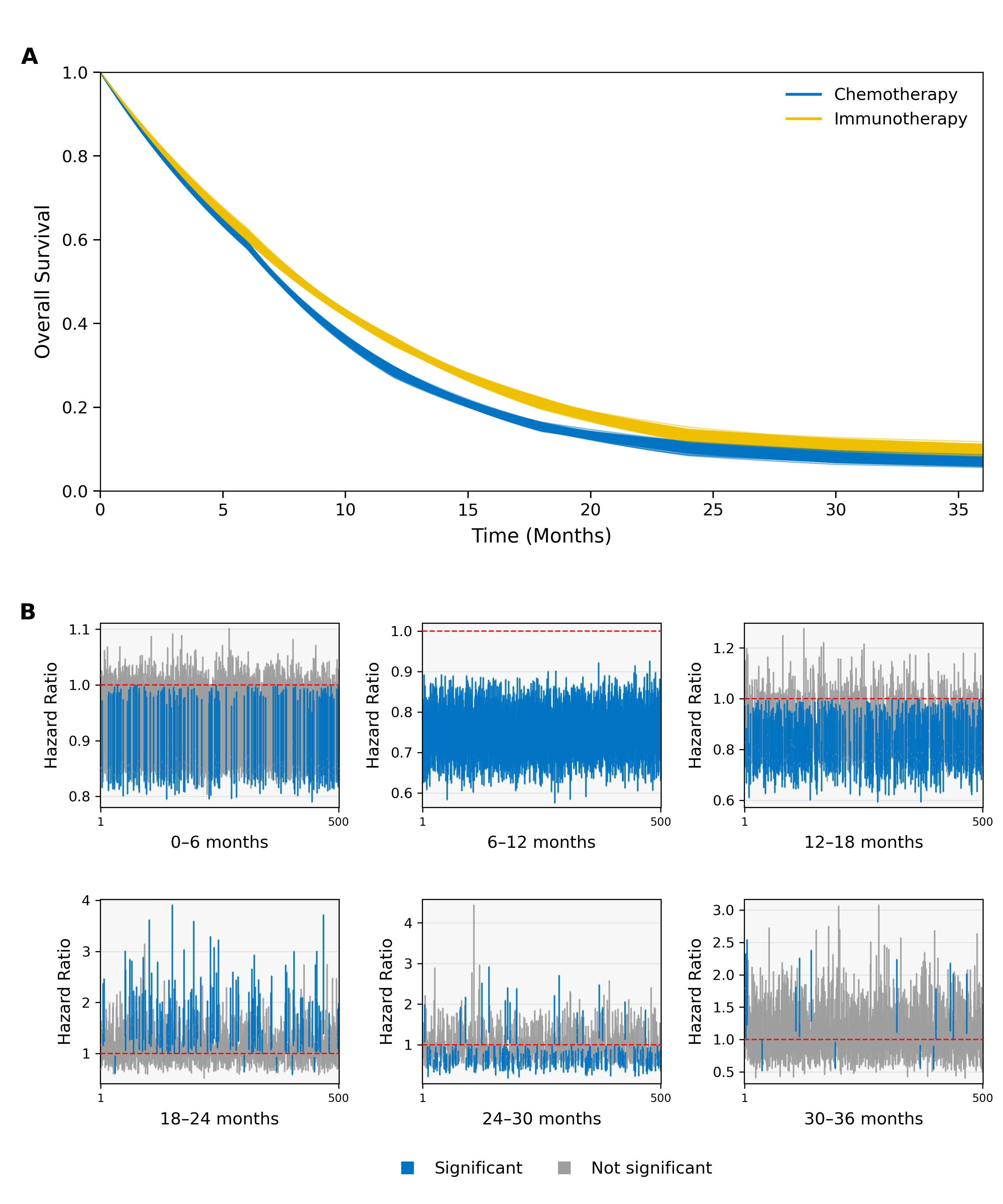}
    \caption{Uncertainty-propagating meta-analysis of immunotherapy (IO) versus chemotherapy (Chemo) in the PD-L1-Low subgroup. (A) Ensemble of OS curves from 500 Monte Carlo simulations of the propagating meta-analysis, demonstrating a consistent separation between IO and Chemo across all plausible subgroup data configurations. (B) Time-varying HR with 95\% CI from each run. Intervals excluding the null value (HR=1) are colored blue, indicating statistical significance.}
    \label{fig_meta_res}
\end{figure}

The pooled survival estimates demonstrate a consistent benefit for immunotherapy. The KM curves from all 500 Monte Carlo simulations form a narrow band of trajectories for each treatment arm, showing clear separation between immunotherapy and chemotherapy across the entire range of plausible subgroup assignments (Fig~\ref{fig_meta_res}A and Supplementary Table~\ref{tab_meta_median_mc_summary}). The pooled mOS for immunotherapy was 8.25 months (range across runs: 8.02–8.44), compared to 7.33 months for chemotherapy (range: 7.21–7.47). Additionally, the 95\% CIs for these mOS estimates were highly stable across all runs. For immunotherapy, the lower bound ranged from 5.49 to 5.96 months and the upper bound from 10.42 to 11.05 months. For chemotherapy, the lower bound ranged from 5.13 to 5.44 months and the upper bound from 8.97 to 9.45 months. This minimal variation demonstrates that the survival estimates are robust to the uncertainties in subgroup label assignment.

The time-varying treatment effect reveals a distinct pattern of benefit across different follow-up periods (Fig~\ref{fig_meta_res}B). The most robust and consistent evidence for immunotherapy was observed during the 6 to 12 month interval. In this period, every Monte Carlo simulation generated a favorable HR (mean HR: 0.75; range: 0.66–0.84). Moreover, this result was statistically significant across all simulations, as the upper bound of the 95\% CI consistently remained below 1 (mean upper bound: 0.84; range: 0.74–0.93).  This demonstrates a conclusive survival advantage for immunotherapy during this specific window. In contrast, the treatment effect in the other intervals, particularly the early (0 to 6 months) and later (18 to 36 months) phases, showed greater statistical uncertainty, with the CIs frequently crossing the null value of 1, indicating a less consistent and non-significant effect (Supplementary Table~\ref{tab_meta_hr_mc_summary}).
 
\section*{Discussion}
The RESOLVE-IPD framework establishes a new paradigm for deriving reliable evidence from published oncology trials by systematically addressing the two most significant barriers in secondary analysis: low-fidelity data extraction and the statistical intractability of subgroup inference. By integrating vector-graphic precision with a novel, uncertainty-aware algorithm, our method moves beyond merely reconstructing data to rigorously quantifying the confidence we can have in the resulting clinical conclusions. This is particularly critical in precision oncology, where treatment decisions often depend on biomarker-defined subgroups that are inconsistently or insufficiently characterized in trial publications.

Our work resolves a foundational flaw in IPD reconstruction that has persisted in existing methodologies. The standard assumption of uniform censoring is a statistical convenience that does not reflect clinical reality, where patient follow-up occurs at discrete intervals. By combining VEC-KM's precise extraction of censoring marks with the CEN-KM algorithm's ability to utilize them, we eliminate this source of systematic bias. The clinical implication is direct: endpoints like the HR and RMST derived from our reconstructed IPD are fundamentally more trustworthy, providing a more solid foundation for meta-analyses that inform clinical decision-making.

A second major innovation is the MAPLE algorithm, which transforms the problem of subgroup data incompleteness into a quantifiable uncertainty. Faced with the common scenario where only aggregate statistics are reported for subgroups, traditional methods force a single, potentially misleading answer. In contrast, MAPLE acknowledges the non-identifiability inherent in the problem and characterizes the entire space of valid solutions. The propagating meta-analysis we demonstrate does not hide this uncertainty; it proactively incorporates it into the final pooled estimate. For the clinical and research community, this means that a conclusion of ``significant benefit" comes with a transparent measure of its robustness, empowering more informed interpretation of the evidence.

We acknowledge several limitations that guide our future development. First, while the high-fidelity of VEC-KM is optimal, its performance depends on the availability of SVG-format figures, which may not always be provided in trial publications. To ensure broad applicability across all publication formats, we will integrate advanced computer vision tools to reliably extract plot coordinates and censoring marks directly from low-fidelity raster images. This enhancement will allow the CEN-KM algorithm to utilize explicit censoring information universally, preserving its accuracy advantage over methods dependent on the uniform censoring assumption. Second, the current implementation of VEC-KM requires manual identification of the figure panels containing the target KM plots and risk tables. We plan to develop a fully automated pipeline for this identification step, which will significantly improve throughput and reproducibility. Finally, the iterative optimization process within the MAPLE module carries a nontrivial computational burden. We will focus on optimizing its efficiency, potentially through machine learning-based surrogates, to accelerate the exploration of the subgroup labeling solution space without compromising the rigorous quantification of uncertainty.

In conclusion, RESOLVE-IPD represents a statistically robust and clinically essential framework for advancing evidence synthesis in oncology. By enabling high-stakes meta-analyses to produce accurate results while transparently addressing inherent uncertainties, this work directly supports the advancement of reliable, precision-focused cancer care.

\section*{Acknowledgment}
Research reported in this publication was supported by the National Institutes of Health under grants R01MH128085 and R01AI197147.

\newpage

\section*{Supplementary Materials of ``RESOLVE-IPD: High-Fidelity Individual Patient Data
Reconstruction and Uncertainty-Aware Subgroup
Meta-Analysis"}

\setcounter{figure}{0}
\setcounter{table}{0}
\setcounter{section}{0}
\setcounter{subsection}{0}

\renewcommand{\thefigure}{S\arabic{figure}}
\renewcommand{\thetable}{S\arabic{table}}
\renewcommand{\thealgorithm}{S\arabic{algorithm}}
\renewcommand{\thesection}{S\arabic{section}}
\renewcommand{\thesubsection}{S\arabic{section}.\arabic{subsection}}

\section{Detailed Methodology for High-Fidelity IPD Reconstruction}
\label{sec_HF_IPD_detail}
This section presents the detailed algorithmic methodology for the VEC-KM (Vector Kaplan-Meier Curve Extraction) and CEN-KM (Censoring-informed Kaplan-Meier Algorithm) modules. These modules handle the digitization of published KM plots and the reconstruction of IPD, respectively, and together form the foundation for accurate IPD recovery. 

\subsection{Detailed Methodology for VEC-KM}
Accurate digitization of KM survival curves is a critical prerequisite for reliable reconstruction of IPD. Errors introduced during digitization can propagate and affect downstream analyses, such as treatment effect estimation and meta-analysis. Common workflows rely on manual point extraction using tools such as DigitizeIt \cite{DigitizeIt} or WebPlotDigitizer \cite{WebPlotDigitizer}, where users align axes and click along the curve to obtain time–survival coordinates. Although accessible, these methods are limited by operator precision and image resolution. Automated image-based approaches (e.g., contour detection and edge tracing) reduce manual workload but remain constrained by figure quality and rasterization artifacts \cite{Zhang2024, ZhaoSunDingXu2025_supp}. In practice, hybrid strategies that combine manual anchors with interpolation are often used, but they can introduce resolution-dependent precision loss, limited reproducibility, and cumulative error over time.
 
To mitigate these limitations, VEC-KM adopts a PDF-native digitization strategy that extracts vector graphics directly from eligible publication files. Unlike image-based methods, vector content preserves the exact shape of KM curves, because each step is encoded as mathematically defined line elements, typically pairs of coordinates marking the start and end of each segment.  This approach faithfully reflects the original information encoded in the published figure, as the extracted coordinates correspond exactly to the curve geometry defined by the authors, rather than an approximation derived from pixel-level interpretation.

VEC-KM proceeds in three major steps: (1) KM curve extraction via connected-segment traversal; (2) axis identification using geometric filtering of point sets; (3) censor mark detection and assignment to curves, following by a final coordinate transformation from page space to time–survival space. These steps progressively reconstruct both the curve geometry and the underlying coordinate system of the published KM plot. In the following sections, we detail each step, including the mathematical formulation and implementation considerations.

\paragraph{KM Curve Extraction} 
We start by extracting all vectorized line segments directly from the PDF using the \textit{get\_drawings} function from python package PyMuPDF~\cite{PyMuPDF}. From the drawing elements, we specifically select graphical primitives of type ``\texttt{l}" (Supplementary Figure~\ref{fig_example_line_seg}), representing straight line segments defined by their start and end coordinates in page space.  For each segment, we record two endpoints \((x_{\mathrm{start}}, y_{\mathrm{start}})\) and \((x_{\mathrm{end}}, y_{\mathrm{end}})\). 

\begin{figure}[htbp!]
\centering
\begin{tcolorbox}[title={Line ("l") Element from SVG Plots},
                  colback=gray!5,
                  colframe=black!30,
                  sharp corners,
                  boxrule=0.5pt,
                  width=0.95\textwidth]
\begin{lstlisting}
{
  "items": [
    ("l",
      Point(36.850399017333984, 102.797119140625),
      Point(566.9293823242188, 102.797119140625)
    )
  ],
  "type": "s",
  "stroke_opacity": 1.0,
  "color": (0.7010, 0.0310, 0.2210),
  "width": 1.5,
  "lineCap": (0, 0, 0),
  "lineJoin": 0.0,
  "closePath": false,
  "dashes": "[] 0",
  "rect": Rect(36.8504, 102.7971, 566.9294, 102.7971),
  "layer": "",
  "seqno": 2,
  "fill": null,
  "fill_opacity": null,
  "even_odd": null
}
\end{lstlisting}
\end{tcolorbox}
\caption{Example of a line segment extracted from a KM curve.}
\label{fig_example_line_seg}
\end{figure}

\begin{figure}[htbp!]
    \centering
    \includegraphics[width=0.95\linewidth]{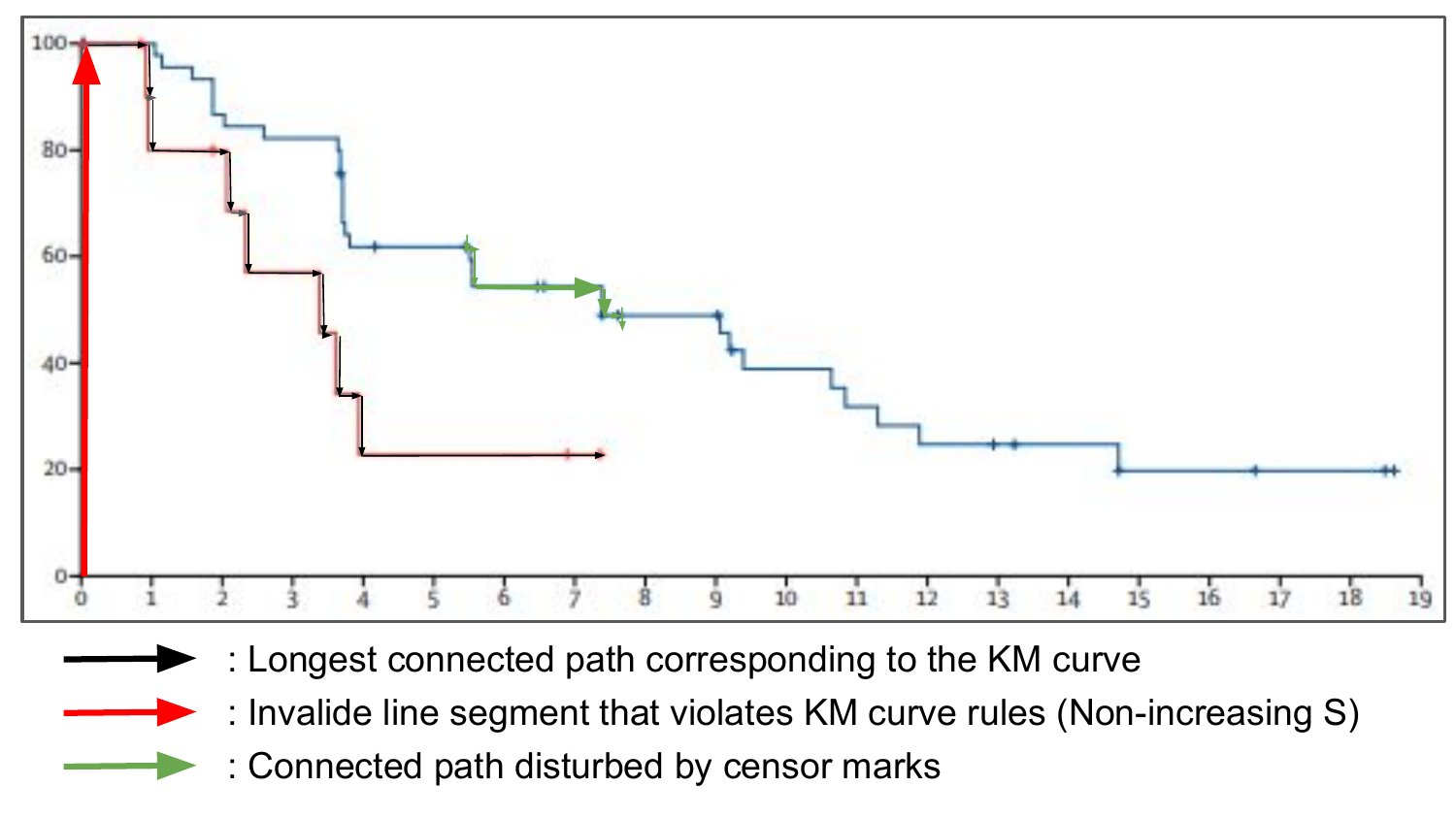}
    \caption{Example of KM curve extraction as a connected path.}
    \label{fig_VEC-KM-Example}
\end{figure}

We then reconstruct continuous curves by connecting segments through endpoint matching, iteratively joining segments where endpoints overlap. This process builds continuous step-shaped trajectories that follow the original KM curve's trajectory. Since KM curves exhibit monotonic behavior: moving forward in time and downward in survival probability, we retain only connected paths that respect this characteristic shape. From the resulting curves, we select the longest paths based on the empirical observation that true KM curves typically comprise the largest number of line segments (Supplementary Figure~\ref{fig_VEC-KM-Example}). The number of selected curves corresponds to the expected number of KM curves reported in the original publication. These reconstructed curves provide the foundation for subsequent analysis steps.

\paragraph{Axis Identification}
To identify the x-axis, we locate pairs of extracted points lying on the same horizontal line and select candidate pairs whose lengths approximate the horizontal span of the KM curve. Among these candidates, we choose the pair positioned closest to the curve as the definitive x-axis. The y-axis is identified through an analogous procedure, selecting vertical point pairs that match the curve's vertical extent. This geometric filtering approach ensures robust axis localization for accurate coordinate transformation.



\paragraph{Censor Mark Detection}
After extracting KM curves, we identify censor marks by examining line segments not incorporated into any reconstructed curve. We apply shape-based rules to detect potential censor marks, focusing on short vertical or cross-shaped line segments with lengths characteristic of typical censoring marks in KM plots. For qualifying segments, we compute the geometric center to establish a stable, resolution-independent representation of the censoring location.

We then classify each candidate censor mark by measuring the vertical distance between its center point and the nearest KM curve. Segments falling within a predefined tolerance threshold are assigned to the corresponding curve. The censoring position is refined by projecting the midpoint onto the associated KM curve, with the horizontal coordinate of this projected point recorded as the censoring time. For overlapping curves where distance-based classification proves ambiguous, we leverage color information from the original figure to resolve assignments. This approach ensures robust recovery of censoring information while accommodating rendering variations in source figures.


\paragraph{Coordinate Transformation.}
The final stage applies a linear transformation to convert digitized points  \((x, y)\) from page coordinates to clinical coordinates \((t, s)\) representing time and survival probability:
\[
t = (x - x_{\text{start}}) \cdot \frac{T_{\text{max}}}{x_{\text{end}} - x_{\text{start}}},
\qquad
s = (y - y_{\text{start}}) \cdot \frac{S_{\text{max}}}{y_{\text{start}} - y_{\text{end}}}.
\]
Here, \(x_{\text{start}}\) and \(x_{\text{end}}\) denote the horizontal positions of the x-axis origin and endpoint in page coordinates, while \(y_{\text{start}}\) and \(y_{\text{end}}\) represent the corresponding vertical positions for the y-axis. \(T_{\text{max}}\) is the maximum reported time on the x-axis, and \(S_{\text{max}}\) is the maximum survival probability on the y-axis (typically equal to 1 or 100). This transformation ensures the recovered time and survival probabilities accurately reflect the original plot's scale and clinical meaning.

\subsection{Detailed Methodology for CEN-KM}
We introduce CEN-KM, a novel algorithm for reconstructing IPD from digitized KM curves. Unlike existing approaches that assume uniform censoring in each time interval, CEN-KM achieves higher accuracy through incorporating explicit censoring marks and matching the number-at-risk table when available.  

\subsubsection{Iterative Event and Censoring Reconstruction}
The algorithm begins with the total number of patients at baseline, $n$, and processes each event time ${t_i}$ and corresponding survival probability $S_i$ extracted from the KM curve sequentially. For each interval $(t_{i-1}, t_i]$, the algorithm first identifies candidate censoring times $c_i \in C$ from available digitized censoring information and inserts them as deterministic data points. The event estimation procedure employs an iterative approach: starting from zero events at time $t_i$, we incrementally add events while computing the KM estimate $\hat{S_i}$ at each step. The process continues until identifying the first event count $m$ that satisfies $\hat{S_i} \leq S_i$. This provides an initial estimate of the number of events needed to produce the observed survival probability drop.

Since each censoring mark may represent multiple censored observations, we implement a multi-branch refinement strategy to address uncertainties arising from potential censoring overlap. We generate three reconstruction branches with $m-1$, $m$, and $m+1$ events at time $t_i$ (omitting the $m-1$ branch when $m=1$). This branching approach captures subtle variations in survival estimates that may result from censoring ambiguities, testing all plausible event counts consistent with the observed survival drop pattern while accounting for the fact that a single graphical censoring mark can correspond to multiple actual censoring events in the underlying data.

For each branch, the algorithm systematically increases the number of overlapping censoring times sampled from the candidate pool $c_i$. This continues until either reaching a predefined maximum number of overlapped censoring times $C_{\max}$ or achieving an absolute difference $|\hat{S_i} - S_i|$ below a specified tolerance threshold. A key advantage of our approach is sampling censoring times from observed candidate positions rather than generating them randomly, which preserves the authentic censoring structure of the original trial and ensures closer alignment with both the published KM curve and risk table.

The branch yielding the smallest absolute deviation from $S_i$ at $t_i$ is selected, completing the reconstruction for that time point. This comprehensive process iterates across all event times. 

\subsubsection{Risk Table Alignment via Censoring Adjustment}

When a number-at-risk table is available, CEN-KM implements a subsequent alignment step that adjusts censoring times at pre-specified boundaries while strictly preserving all event times. This ensures precise agreement between the reconstructed IPD and both the published survival curve and risk table, without altering the fundamental shape of the KM survival function.

The alignment process iterates through each boundary time point $\tau$ specified in the risk table. For each $\tau$, we compute the current number of patients at risk, defined as those with event or censoring times $\geq \tau$, and compare this to the target value from the risk table. The discrepancy is calculated as $\Delta = \text{target} - \text{current}$. If $\Delta = 0$, the reconstructed IPD already matches the risk table at this boundary and no adjustment is required. When $\Delta > 0$, indicating insufficient patients at risk, the algorithm identifies censoring times occurring before $\tau$ and removes them. This effectively extends patient follow-up duration, increasing the number of individuals considered at risk at time $\tau$. To maintain the integrity of the censoring distribution, overlapped censors are prioritized for removal. Conversely, if $\Delta < 0$, indicating excess patients at risk, the algorithm introduces additional censoring times sampled from the candidate pool $c_i$ before $\tau$ to reduce the at-risk count appropriately.

\begin{figure}[htbp!]
    \centering
    \includegraphics[width=0.95\linewidth]{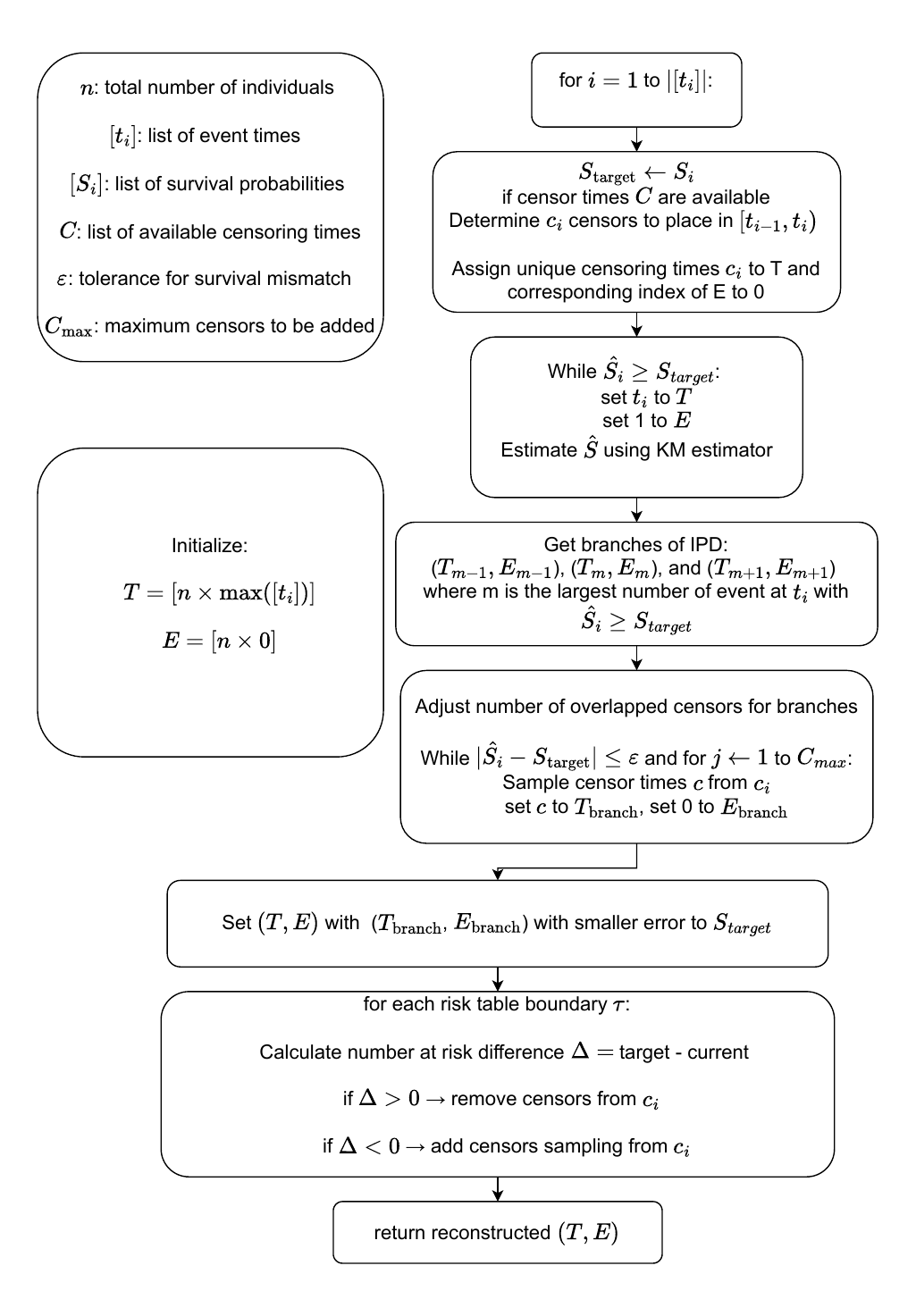}
    \caption{CEN-KM Algorithm}
    \label{fig_ReconIPD_Algo}
\end{figure}

This local adjustment procedure repeats sequentially across all time points in the number-at-risk table. The final result is a globally coherent IPD reconstruction that simultaneously matches both the KM survival curve and the reported numbers at risk at each time point. Critically, because only censoring times are modified while event times remain unchanged, the algorithm preserves the exact survival probability drops observed in the published KM curve, ensuring consistency with the original survival estimates. Supplementary Figure \ref{fig_ReconIPD_Algo} summarizes the CEN-KM algorithm.

\subsubsection{Methodological Contributions}

CEN-KM introduces three key methodological advances over existing approaches:
\begin{itemize}
\item \textbf{Multi-branch event estimation}: The branching strategy at each event time captures reconstruction uncertainties arising from censoring overlap, providing more robust event count estimation.

\item \textbf{Observed censoring preservation}: By sampling from actual candidate censoring positions rather than generating synthetic times, CEN-KM maintains the trial's censoring pattern.

\item \textbf{Event-preserving risk table alignment}: The targeted boundary adjustment modifies only censoring times, ensuring exact KM curve reproduction while achieving risk table concordance.
\end{itemize}

\section{Detailed Methodology for Uncertainty-Aware Subgroup IPD Recovery}

\subsection{KM Subtraction in Deterministic Pathway}
\label{sec_kmsub_detail}
When high-fidelity KM curves are available for both the overall cohort and all but one subgroup, the missing subgroup's IPD can be reconstructed using the KM subtraction procedure~\cite{zhao2022kmsubtraction_supp}. This approach leverages the mathematical relationship between overall population and subgroup survival curves through an optimal matching process.

The method begins by reconstructing IPD from both the overall population KM curve and the known subgroup's KM curve. Patients are categorized into event and censoring groups within each dataset. The procedure involves formulating and solving an optimal matching problem between patients from the overall population and the known subgroup based on their follow-up times.
This constitutes a bipartite matching problem where we compute pairwise distance matrices between patients, typically using absolute or squared differences in follow-up times. The matching algorithm then identifies the optimal assignment that minimizes the total matching cost across all patient pairs. Common approaches include the Hungarian algorithm for minimum-cost perfect matching, nearest-neighbor matching for greedy pairing of closest patients, and Mahalanobis distance matching when additional covariates are available for incorporation. After solving the matching optimization, successfully paired patients are classified as belonging to the known subgroup, while the remaining unpaired patients from the overall population are identified as members of the missing subgroup. This procedure extends naturally to settings with multiple known subgroups through sequential application of KM subtraction, progressively matching the overall IPD against each known subgroup to isolate the unreported subgroup's patients.

\subsection{MAPLE (Marginal Assignment of Plausible Subgroup Labels and Evidence Propagation)} 
\label{sec_maple_detail}
When subgroup-level KM curves are unavailable, RESOLVE-IPD employs its optimization-based MAPLE pathway to infer subgroup labels. This approach leverages the subgroup summary statistics commonly reported alongside mOS results for overall population in clinical trial publications. For example, as shown in Supplementary Figure~\ref{fig_attraction_forest_plot}, the ATTRACTION-3 trial evaluating Nivolumab versus chemotherapy in advanced oesophageal squamous cell carcinoma reported OS curves along with forest plots containing subgroup analyses~\cite{kato2019nivolumab_supp}. 
These forest plots typically include subgroup sample sizes, event counts, and HRs with confidence intervals. Some publications additionally report subgroup-specific mOS with confidence intervals. These reported statistics enable the estimation of plausible subgroup assignments through constrained optimization. Specifically, MAPLE aims to assign subgroup labels to patients in the reconstructed OS dataset of overall population such that the resulting stratified data reproduces both the exact subgroup counts and the reported statistical targets. 

 \begin{figure}[!htbp]
    \centering
    \includegraphics[width=0.9\linewidth]{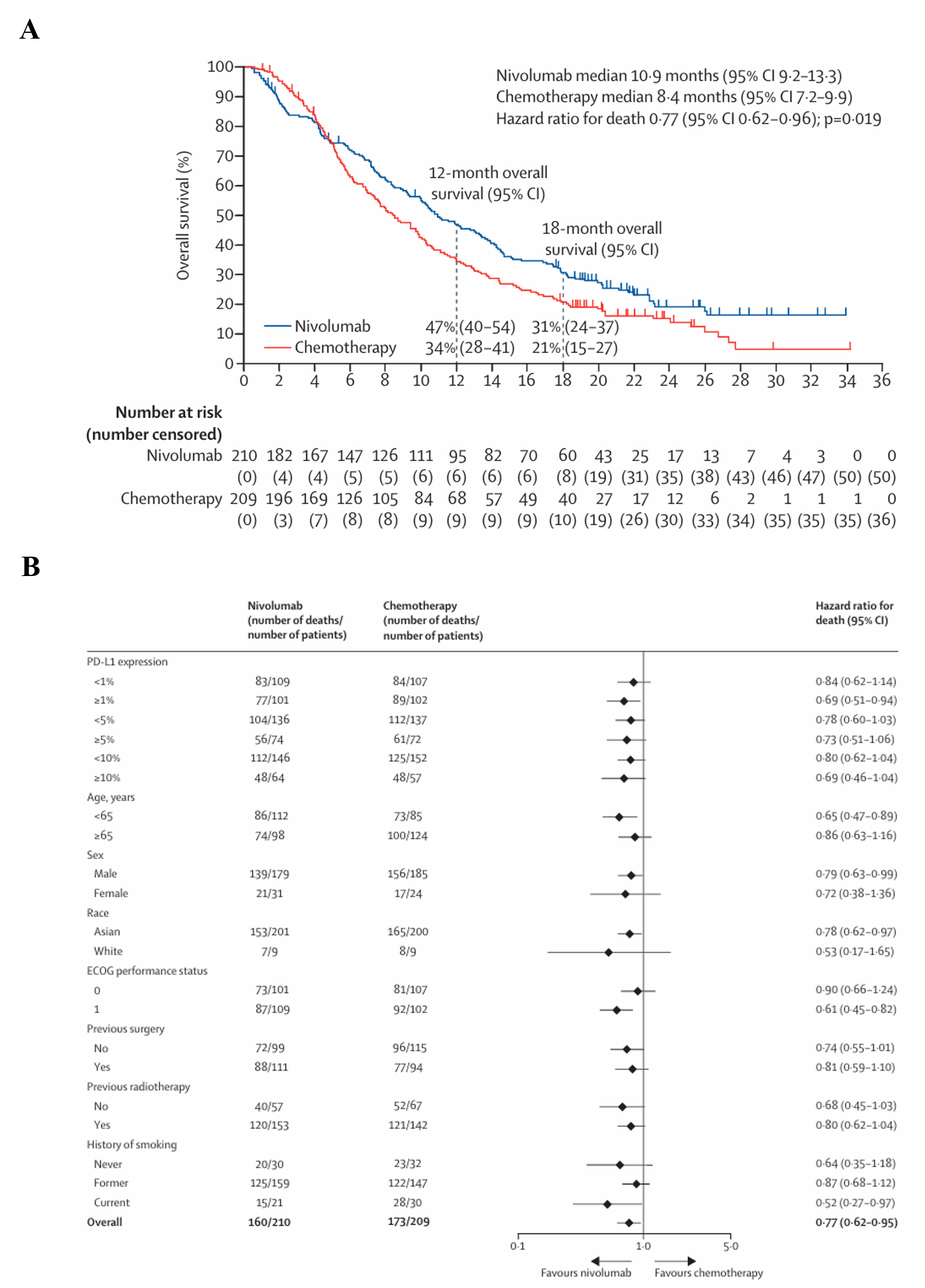}
    \caption{(A) OS curve from the original ATTRACTION-3 publication (B) Forest plot from the original ATTRACTION-3 publication}
    \label{fig_attraction_forest_plot}
\end{figure}

\subsubsection{Problem Setup and Notation}
Let $\widetilde{\mathcal{D}} = {(\tilde{t}_i, \tilde{\delta}_i, \tilde{Z}_i): i = 1, \dots, n}$ denote the reconstructed IPD obtained from digitizing the overall trial-level KM curves, where $\tilde{t}_i$ is the reconstructed survival time, $\tilde{\delta}_i \in \{0,1\}$ is the event indicator (1 for event, 0 for censored), and $\tilde{Z}_i \in \{0,1\}$ indicates treatment assignment. Without loss of generality, we consider two treatment arms in this work; however, the proposed methods can be readily extended to trials with multiple arms. 
The objective is to assign each patient $i$ a subgroup label $g_i \in \{0, 1, \dots, K-1\}$, where $K$ denotes the total number of predefined subgroups (e.g., based on biomarker status, disease stage, or other clinical categories). For example, when inferring PD-L1 expression status subgroups (high vs. low) as in the ATTRACTION-3 trial, $K=2$ with $g_i=0$ representing PD-L1 low and $g_i=1$ representing PD-L1 high.  Let $g = (g_1, \dots, g_n)$ denote a candidate subgroup labeling.

Two types of constraints are typically available from published subgroup analyses: 
\begin{itemize}
\item Hard Count Constraints: These include subgroup sample sizes $n_k$, numbers of observed events $E_k$, and treatment-specific subgroup counts $n_{k0}$ and $n_{k1}$ for each subgroup $k$. These must be satisfied exactly by any feasible labeling $g$. 
\item Statistical Targets with Uncertainty: These generally include subgroup-specific HRs $\mathrm{HR}_k^{\mathrm{tgt}}$ with associated confidence intervals $[\mathrm{HR}_k^{\mathrm{L}}, \mathrm{HR}_k^{\mathrm{U}}]$, and may also include treatment-specific mOS times $\mathrm{Med}_{kZ}^{\mathrm{tgt}}$ for $Z \in \{0,1\}$ with confidence intervals $[\mathrm{Med}_{kZ}^{\mathrm{L}}, \mathrm{Med}_{kZ}^{\mathrm{U}}]$. These statistical targets with confidence intervals should be matched closely.
\end{itemize}

Our goal is to find labelings $g = (g_1,\dots,g_n)$ such that the reconstructed dataset $(\widetilde{\mathcal{D}}, g)$ satisfies the hard count constraints 
and for each subgroup $k$ and treatment arm $Z$: 
\[
\widehat{\mathrm{HR}}_k(g)\approx \mathrm{HR}_k^{\mathrm{tgt}},\qquad
\widehat{\mathrm{Med}}_{kZ}(g)\approx \mathrm{Med}_{kZ}^{\mathrm{tgt}},
\]
\[
\widehat{\mathrm{CI}}(\widehat{\mathrm{HR}}_k(g))\approx [\mathrm{HR}_k^{\mathrm{L}},\,\mathrm{HR}_k^{\mathrm{U}}],\qquad
\widehat{\mathrm{CI}}(\widehat{\mathrm{Med}}_{kZ}(g))\approx [\mathrm{Med}_{kZ}^{\mathrm{L}},\,\mathrm{Med}_{kZ}^{\mathrm{U}}],
\]
where quantities computed from $\widetilde{\mathcal{D}}$ under labeling $g$ are denoted with hats. 

\subsubsection{Optimization Formulation}
We formulate the subgroup label recovery as a constrained optimization problem that identifies assignments $g = (g_1, \dots, g_n)$ satisfying all reported hard constraints while minimizing the maximum relative deviation between recomputed and target summary statistics.

Let $\Gamma(g; \widetilde{\mathcal{D}}) \in \mathbb{R}^m$ denote the vector of hard constraint functions, typically comprising subgroup-wise counts such as sample sizes, event totals, and treatment allocations. Let $\gamma \in \mathbb{R}^m$ represent the corresponding reported values. Feasibility requires:
\[
\Gamma(g;\widetilde{\mathcal{D}})=\gamma \ \ \ \ \ \ \mathrm{(hard \ constraints)}.
\] 

Let $\widehat{\bm{\tau}}(g; \widetilde{\mathcal{D}})$ denote the vector of recomputed subgroup statistics and $\bm{\tau}^{\mathrm{tgt}}$ the corresponding reported targets. For each component $j$, we define the relative deviation as:
\[
r_j\big(g;\widetilde{\mathcal{D}},\bm{\tau}^{\mathrm{tgt}}\big)=\frac{\big|\,\widehat{\tau}_j(g;\widetilde{\mathcal{D}})-\tau^{\mathrm{tgt}}_j\,\big|}{\big|\,\tau^{\mathrm{tgt}}_j\,\big|}.
\]
The optimization problem is then formulated as:
\[
\min_{g}\max_jr_j\big(g;\widetilde{\mathcal{D}}, \bm{\tau}^{\mathrm{tgt}}\big) \ \ \ \ \text{subject to}\quad \Gamma(g;\widetilde{\mathcal{D}})=\gamma.
\]

This $\ell_\infty$ relative loss function is scale-free and intentionally strict, controlling the worst-case fractional error across all matched summaries. By preventing any single metric from substantially deviating from its target, this formulation promotes precise recovery of all reported statistics. The optimization thus identifies labelings $g$ that exactly satisfy the hard constraints $\gamma$ while minimizing the maximum relative deviation between the recomputed statistics $\widehat{\bm{\tau}}(g; \widetilde{\mathcal{D}})$ and the reported targets $\bm{\tau}^{\mathrm{tgt}}$.

\subsubsection{Algorithmic Implementation via Simulated Annealing}
This optimization problem is inherently challenging. The discrete subgroup labels create a combinatorial search space, while stringent count constraints on subgroup sizes, event totals, and treatment allocations introduce complex coupling between variables. The objective function exhibits particular difficulty due to the nonlinear survival estimators that create a highly nonconvex and nonsmooth optimization landscape. The $\ell_\infty$ relative loss further complicates these challenges by creating sharp ridges that emphasize the single worst-matching component. Compounding these issues, the problem suffers from inherent non-identifiability: recovering high-dimensional subgroup labels from low-dimensional summary statistics, combined with rounding in published targets, means multiple distinct labelings may achieve similar objective values. Collectively, these characteristics define a large-scale constrained combinatorial optimization problem analogous to mixed-integer nonlinear programming, for which exact methods become computationally intractable at realistic sample sizes.


Given this challenging landscape, we employ metaheuristic optimization approaches~\cite{parejo2012metaheuristic} that balance effective local search with sustained global exploration. We implement simulated annealing (SA) ~\cite{kirkpatrick1983optimization} with a specialized neighborhood structure that preserves feasibility throughout the search process. The algorithm begins with a feasible initial labeling satisfying all hard constraints $\Gamma(g;\widetilde{\mathcal{D}})=\gamma$ and employs balanced same-cell swaps as neighborhood moves. These swaps exchange subgroup labels between two patients matched on treatment assignment and event status, ensuring all hard constraints remain satisfied automatically. Simulated annealing proves particularly suitable for this problem due to its ability to handle rugged, nonsmooth objectives; its occasional acceptance of uphill moves facilitates escape from local optima and traversal of rounding-induced plateaus; and its geometric cooling schedule provides a straightforward mechanism for balancing exploration and exploitation with minimal hyperparameter tuning.  The complete simulated annealing procedure with balanced same-cell swaps is detailed in Algorithm~\ref{alg_sa_balanced_swap_compact}. 

To enhance exploration of the solution space and improve computational efficiency, we execute multiple independent simulated annealing runs from diverse feasible initializations. Specifically, for seeds $s=1,\ldots,S$, we launch parallel runs with distinct random initializations, each satisfying the hard constraints $\Gamma(g^{(0)})=\gamma$. This parallelization yields a collection of $S$ candidate solutions ${(\widehat{L}_s, \widehat{g}_s)}_{s=1}^S$ that collectively represent diverse regions of the feasible space. These solutions subsequently undergo filtering and processing within the broader MAPLE framework to address the inherent non-identifiability and generate the final ensemble of plausible subgroup labelings.

\begin{algorithm}[htbp!]
\caption{Simulated annealing with balanced same-cell swaps}
\label{alg_sa_balanced_swap_compact}
\small

\textbf{Inputs:} reconstructed IPD $\widetilde{\mathcal{D}}$; targets $\bm{\tau}^{\mathrm{tgt}}$; hard constraints $\gamma$; loss function $f_{\mathrm{loss}}(g;\widetilde{\mathcal{D}},\bm{\tau}^{\mathrm{tgt}})\in[0,\infty)$;\\
initial temperature $T_0>0$; cooling factor $\rho\in(0,1)$; iteration budget $\textit{max\_iter}\in\mathbb{N}$;\\
minimum iteration threshold $\textit{min\_iter}\in\mathbb{N}_0$; stagnation limit $\textit{stagnation\_stop\_iter}\in\mathbb{N}$.

\medskip
\textbf{Initialization:}
\begin{enumerate}
\item Randomly pick $g^{(0)}$ such that $\Gamma(g^{(0)};\widetilde{\mathcal{D}})=\gamma$.
\item Set 
\[
T \leftarrow T_0, \quad
L^{(0)} \leftarrow f_{\mathrm{loss}}\big(g^{(0)};\widetilde{\mathcal{D}},\bm{\tau}^{\mathrm{tgt}}\big), \quad
(\widehat{g},\widehat{L}) \leftarrow (g^{(0)},L^{(0)}), \quad
s \leftarrow 0.
\]
\end{enumerate}

\medskip
\textbf{Main loop ($t=0,1,\ldots,\textit{max\_iter}-1$):}
\begin{enumerate}
\item \textit{Balanced same-cell swap:}\\
Randomly pick $(i,j)$ such that $(\tilde Z_i,\tilde\delta_i)=(\tilde Z_j,\tilde\delta_j)$ and $g^{(t)}_i\neq g^{(t)}_j$; define
\[
g' \leftarrow g^{(t)} \text{ with } (g'_i,g'_j)=(g^{(t)}_j,g^{(t)}_i).
\]

\item \textit{Evaluate and accept (Metropolis rule):}
\[
\begin{aligned}
L^{(t)} &= f_{\mathrm{loss}}\big(g^{(t)};\widetilde{\mathcal{D}},\bm{\tau}^{\mathrm{tgt}}\big),\quad
L'      = f_{\mathrm{loss}}\big(g';\widetilde{\mathcal{D}},\bm{\tau}^{\mathrm{tgt}}\big),\\
\Delta  &= L' - L^{(t)}, \qquad
a(T,\Delta) = 
\begin{cases}
1, & \Delta \le 0,\\
\exp(-\Delta/T), & \Delta>0.
\end{cases}
\end{aligned}
\]
Draw $U\sim\mathrm{Unif}[0,1]$ and set
\[
(g^{(t+1)},L^{(t+1)}) =
\begin{cases}
(g',L'), & U \le a(T,\Delta),\\
(g^{(t)},L^{(t)}), & \text{otherwise.}
\end{cases}
\]

\item \textit{Best-so-far update and cooling:}
\[
\text{if } L^{(t+1)}<\widehat{L} :
(\widehat{g},\widehat{L})\!\leftarrow\!(g^{(t+1)},L^{(t+1)}),\, s\!\leftarrow\!0;
\quad
\text{else } s\!\leftarrow\!s\!+\!1;
\quad
T\!\leftarrow\!\rho\,T.
\]

\item \textit{Stopping check:}
\[
\text{if } 
\widehat{L}=0 
\text{ or } 
(t+1 \ge \textit{min\_iter} \text{ and } s \ge \textit{stagnation\_stop\_iter}),
\textbf{ stop and return } (\widehat{g},\widehat{L}).
\]
\end{enumerate}

\medskip
\textbf{Outputs:} best loss $\widehat{L}$, best labeling $\widehat{g}$.
\end{algorithm}

\subsubsection{Addressing Non-Identifiability}

MAPLE addresses the inherent non-identifiability through a one- or two-stage ensemble generation process that produces multiple plausible subgroup labelings:

\begin{itemize}
\item \textbf{Initial candidate generation}: We first identify the minimal loss value among all S candidate solutions from the optimization, and collect all labelings that achieve this minimum, denoted as $\mathcal{G}_{\mathrm{MAPLE}}$. If no additional information from the original publication is available to further refine the label assignments, $\mathcal{G}_{\mathrm{MAPLE}}$ constitutes the final output of the MAPLE framework.

\item \textbf{Evidence-based refinement}: When publications provide subgroup KM plots as low-fidelity raster images, we leverage existing digitization tools such as KM-GPT~\cite{ZhaoSunDingXu2025_supp} to reconstruct approximate subgroup-level IPD. We then filter the candidate labelings, retaining only those whose derived subgroup KM curves fall within the 95\% confidence intervals of the reconstructed curves. The resulting collection of labelings is denoted $\mathcal{G}_{\mathrm{MAPLE}}^*$ and serves as the final output of MAPLE.
\end{itemize}

This methodology ensures consistency with all available evidence while explicitly preserving the uncertainty inherent in this non-identifiable inverse problem.
The final output is an ensemble $\mathcal{G}_{\mathrm{MAPLE}}$ or $\mathcal{G}_{\mathrm{MAPLE}}^*$ of plausible subgroup labelings, where each member represents a distinct, data-compatible reconstruction of the subgroup-level IPD. This ensemble provides principled uncertainty quantification for subgroup assignment and serves as the foundation for uncertainty-aware downstream analyses within the RESOLVE-IPD framework.

\section{Synthetic Data Generation}
\label{sec_sim_generation}
We generated right-censored survival data with two latent label groups $g \in \{0,1\}$, representing biomarker high and biomarker low group, and a randomized treatment indicator $Z \in \{0,1\}$. The total sample size was fixed at $N = 400$, with 200 individuals assigned to group $g=0$, and the remaining 200 to group $g=1$. The treatment indicator was independently generated as $Z_i \sim \mathrm{Bernoulli}(0.5)$ for $i=1,\ldots,N$.

Within each label group, survival times followed an exponential distribution with constant baseline hazard $\lambda_0=0.1$ and group-specific treatment effects expressed as HRs. The true HRs were $\mathrm{HR}_{g=0}=0.9$ and $\mathrm{HR}_{g=1}=0.7$, corresponding to log–hazard coefficients $\beta_0=\log(0.9)$ and $\beta_1=\log(0.7)$. For an individual with label $g_i$ and treatment $Z_i$, the event time was generated from an exponential distribution with rate $\lambda_0 \exp(\beta_{g_i} Z_i)$, that is,
\[
T_i \mid (g_i,Z_i) \sim \mathrm{Exponential}\big(\lambda_0 \exp(\beta_{g_i} Z_i)\big).
\]

Independent censoring times were simulated using a two-component mixture of uniform distributions to reflect patterns of administrative and early censoring. Specifically, censoring times (\( C_i \)) were drawn with \( 90\% \) probability from \( \mathrm{Uniform}(12, 24) \) months and \( 10\% \) probability from \( \mathrm{Uniform}(0, 12) \) months. These censoring times were independently generated with respect to the survival times (\( T_i \)), latent label groups (\( g_i \)), and treatment indicators (\( Z_i \)).  The observed survival time was defined as $Y_i=\min(T_i,C_i)$, and the event indicator as $\Delta_i=\mathbb{I}\{T_i \le C_i\}$. The resulting dataset therefore consisted of tuples $(Y_i, \Delta_i, Z_i, g_i)$ for $i=1,\ldots,N$.

\section{Hierarchical Piecewise-Exponential Model for Meta-Analysis}
\label{sec_meta_model}

This section details the Bayesian hierarchical piecewise-exponential model used for meta-analysis. The goal is to synthesize survival data from multiple studies, estimating a pooled hazard function that evolves smoothly over time while formally accounting for between-study heterogeneity.

The time axis is partitioned into $K$ disjoint intervals, $\cT_k = (c_{k-1}, c_k]$ for $k=1,\dots,K$. The hazard function for each study is assumed to be constant within these intervals.
Let $\theta_{sk}$ represent the log-hazard for study $s$ in time interval $k$, so that the study-specific hazard is $h_{sk} = \exp(\theta_{sk})$. To pool information across studies, the $\theta_{sk}$ are modeled hierarchically:

\begin{equation}
\theta_{sk} \mid \beta_k, \nu_k^2 \sim \mathcal{N}(\beta_k, \nu_k^2).
\end{equation}
Here, $\beta_k$ is the meta-analytic average log-hazard in interval $k$, and $\nu_k$ captures the between-study heterogeneity for that interval.

To share information across intervals and ensure the pooled hazard function is smooth, we place a hierarchical prior on the sequence $\bmeta = (\beta_1, \ldots, \beta_K)$. Each $\beta_k$ is modeled as varying around a latent mean $\gamma_k$:

\begin{equation}
\beta_k \mid \gamma_k, \omega^2 \sim \mathcal{N}(\gamma_k, \omega^2),
\end{equation}
where $\omega$ controls the deviation of the $\beta_k$ from the latent process.

The latent process $\bgamma = (\gamma_1, \ldots, \gamma_K)$ is assigned a stationary autoregressive prior of order one (AR(1)) to enforce smoothness:

\begin{align*}
\gamma_1 \mid \rho, \delta^2 &\sim \mathcal{N}\left(0, \frac{\delta^2}{1-\rho^2}\right), \\
\gamma_k \mid \gamma_{k-1}, \rho, \delta^2 &\sim \mathcal{N}\left(\rho \, \gamma_{k-1}, \delta^2\right), \quad \text{for } k=2,\ldots,K,
\end{align*}
with $|\rho| < 1$ ensuring stationarity and $\delta > 0$.

We complete the model with the following prior distributions:

\begin{itemize}
    \item The heterogeneity standard deviations are assigned weakly informative half-Normal priors: $\nu_k \sim \mathcal{N}^+(0, 0.25^2)$ and $\omega \sim \mathcal{N}^+(0, 0.25^2)$.
    \item The standard deviation of the latent process is given a prior: $\delta \sim \mathcal{N}^+(0, 1^2)$.
    \item The autoregressive parameter $\rho$ is constrained to the stationary region $(-1, 1)$ via the transformation $\rho = 2 \times \mathrm{logit}^{-1}(\zeta) - 1$, with the prior $\zeta \sim \mathcal{N}(0, 1)$.
\end{itemize}

The pooled hazard in interval $k$ is $\lambda^{\text{meta}}_k = \exp(\beta_k)$, and the study-specific hazard is $h_{sk} = \exp(\theta_{sk})$. Let $L_k = c_k - c_{k-1}$ be the length of the $k$-th interval and let $k(t)$ be the index of the interval containing time $t$.

The survival probability for the meta-analytic population is:
\begin{align*}
S_{\text{meta}}(t) &= \exp\left( -\sum_{m=1}^{k(t)} \lambda^{\text{meta}}_m L_m \right),
\end{align*}
and for study $s$, it is:
\begin{align*}
S_s(t) &= \exp\left( -\sum_{m=1}^{k(t)} h_{sm} L_m \right).
\end{align*}

\section{Supplementary Figures and Tables}

\begin{figure}[H]
    \centering
    \includegraphics[width=\linewidth]{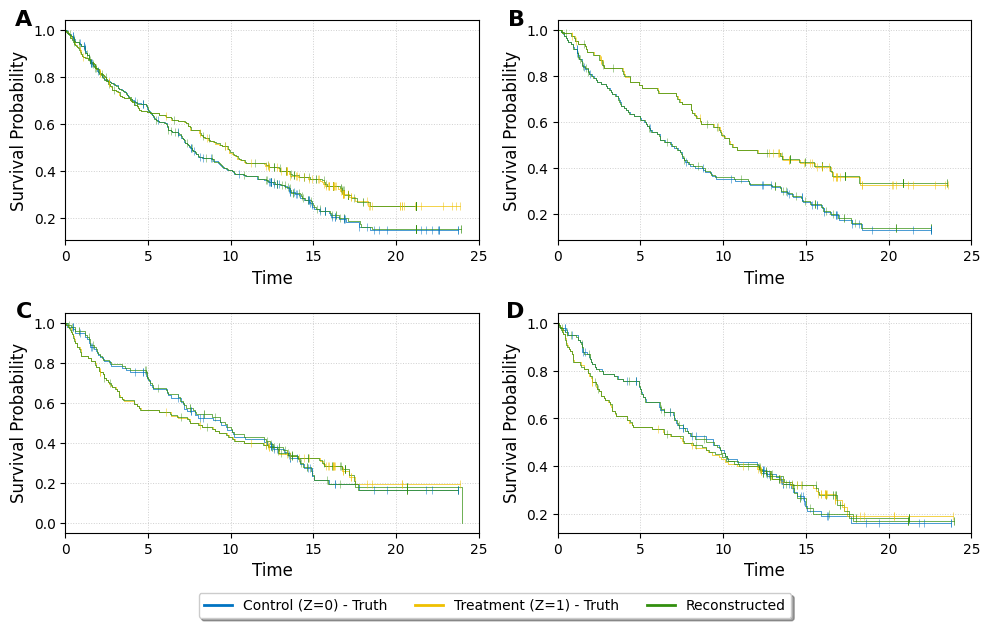}
    \caption{KM plots derived from reconstructed subgroup-level IPD using VEC-KM + IPDfromKM, overlaid with the true KM curves. VEC-KM accurately digitizes the KM figures, enabling IPDfromKM to achieve good performance in reconstructing survival data. However, the censoring patterns in the reconstructed data differ from the true distribution, as they assume uniform censoring within each interval. Subplots depict KM curves for (A) the overall population, (B) the PD-L1–high subgroup, (C) the PD-L1–low subgroup, and (D) the PD-L1–low subgroup obtained via KM subtraction.}
    \label{fig_sim_ipdfromkm}
\end{figure}

\begin{figure}[H]
    \centering
    \includegraphics[width=\linewidth]{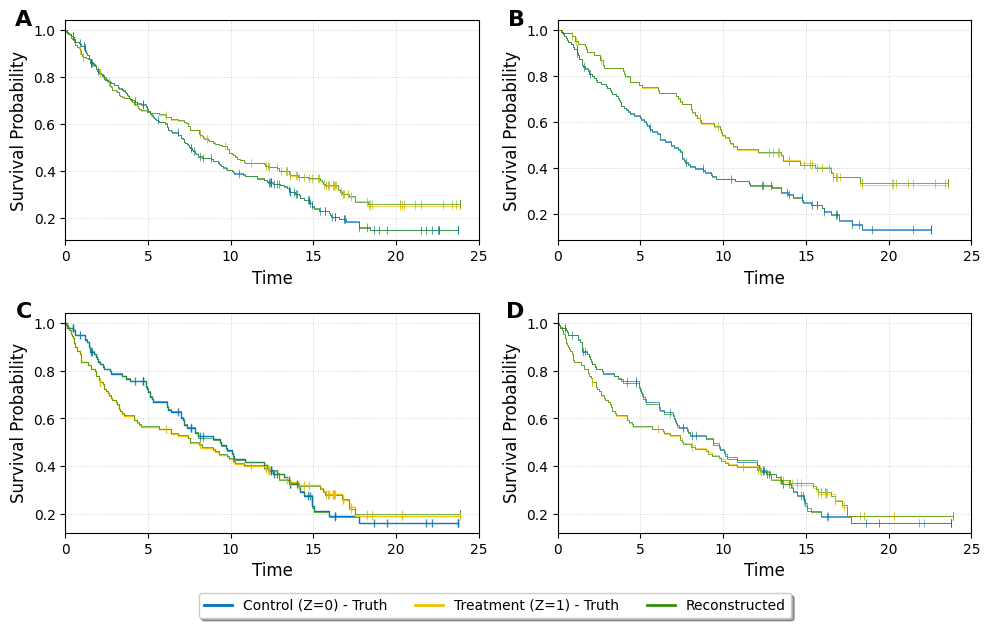}
    \caption{KM plots derived from reconstructed subgroup-level IPD using VEC-KM + KMtoIPD, overlaid with the true KM curves. VEC-KM performs well at accurately digitizing the KM figures, contributing to the strong performance of KMtoIPD in IPD reconstruction. However, KMtoIPD assumes that each censoring mark corresponds to a single censored patient, which introduces errors in the reconstructed IPD. Subplots depict KM curves for (A) the overall population, (B) the PD-L1–high subgroup, (C) the PD-L1–low subgroup, and (D) the PD-L1–low subgroup obtained via KM subtraction.
}
    \label{fig_sim_kmtoipd}
\end{figure}

\begin{figure}[H]
    \centering
    \includegraphics[width=0.8\linewidth]{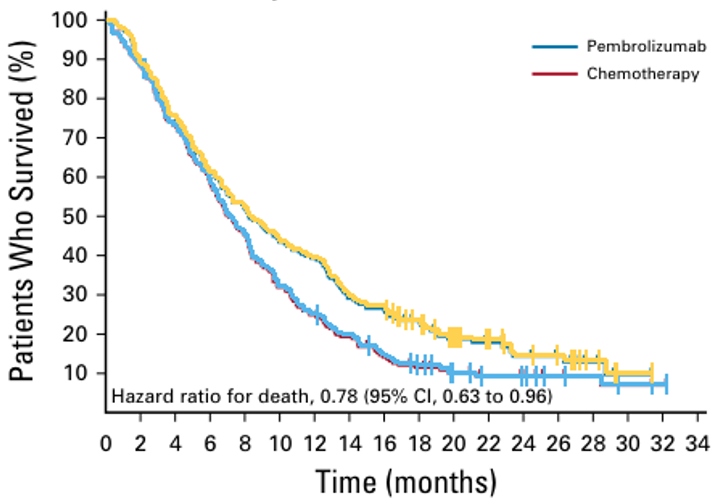}
    \caption{Overlay of reconstructed KM curves on the original plots for the overall population in KEYNOTE-181. The yellow line (IO) and blue line (Chemo) represent the reconstructed curves obtained through individual patient data (IPD) generated using VEC-KM + CEN-KM.}
    \label{fig_meta_keynote_overall_overlay}
\end{figure}

\begin{figure}[H]
    \centering
    \includegraphics[width=\linewidth]{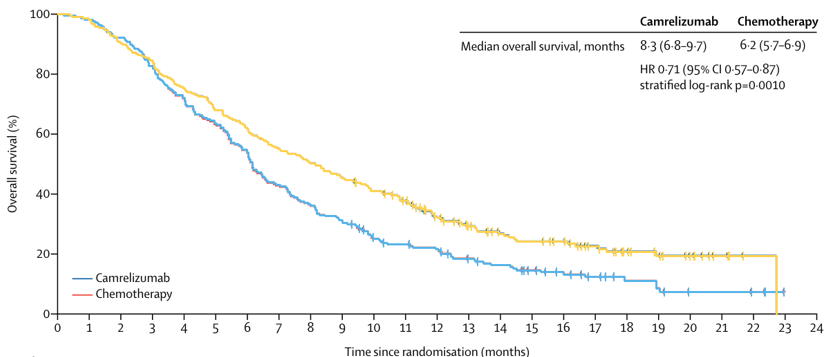}
    \caption{Overlay of reconstructed KM curves on the original plots for the overall population in ESCORT. The yellow line (IO) and blue line (Chemo) represent the reconstructed curves obtained through individual patient data (IPD) generated using VEC-KM + CEN-KM.}
    \label{fig_meta_escort_overall_overlay}
\end{figure}

\begin{figure}[H]
    \centering
    \includegraphics[width=\linewidth]{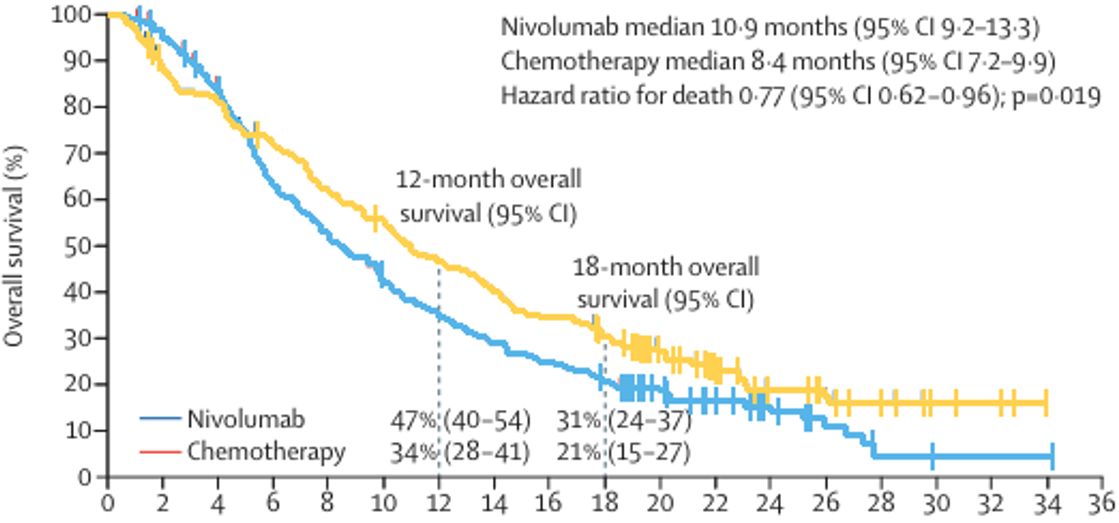}
    \caption{Overlay of reconstructed KM curves on the original plots for the overall population in ATTRACTION-3. The yellow line (IO) and blue line (Chemo) represent the reconstructed curves obtained through individual patient data (IPD) generated using VEC-KM + CEN-KM.}
    \label{fig_meta_attraction_overall_overlay}
\end{figure}

\begin{figure}[H]
    \centering
    \includegraphics[width=\linewidth]{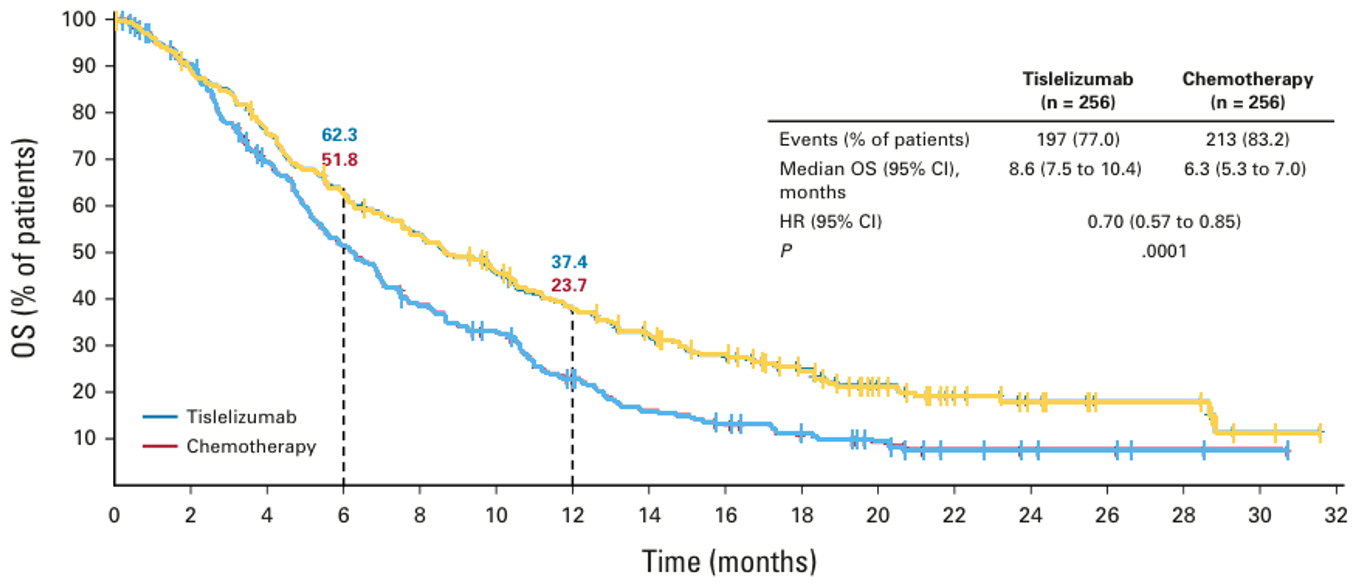}
    \caption{Overlay of reconstructed KM curves on the original plots for the overall population in RATIONALE-302. The yellow line (IO) and blue line (Chemo) represent the reconstructed curves obtained through individual patient data (IPD) generated using VEC-KM + CEN-KM. }
    \label{fig_meta_rationale_overall_overlay}
\end{figure}

\begin{figure}[H]
    \centering
    \includegraphics[width=\textwidth]{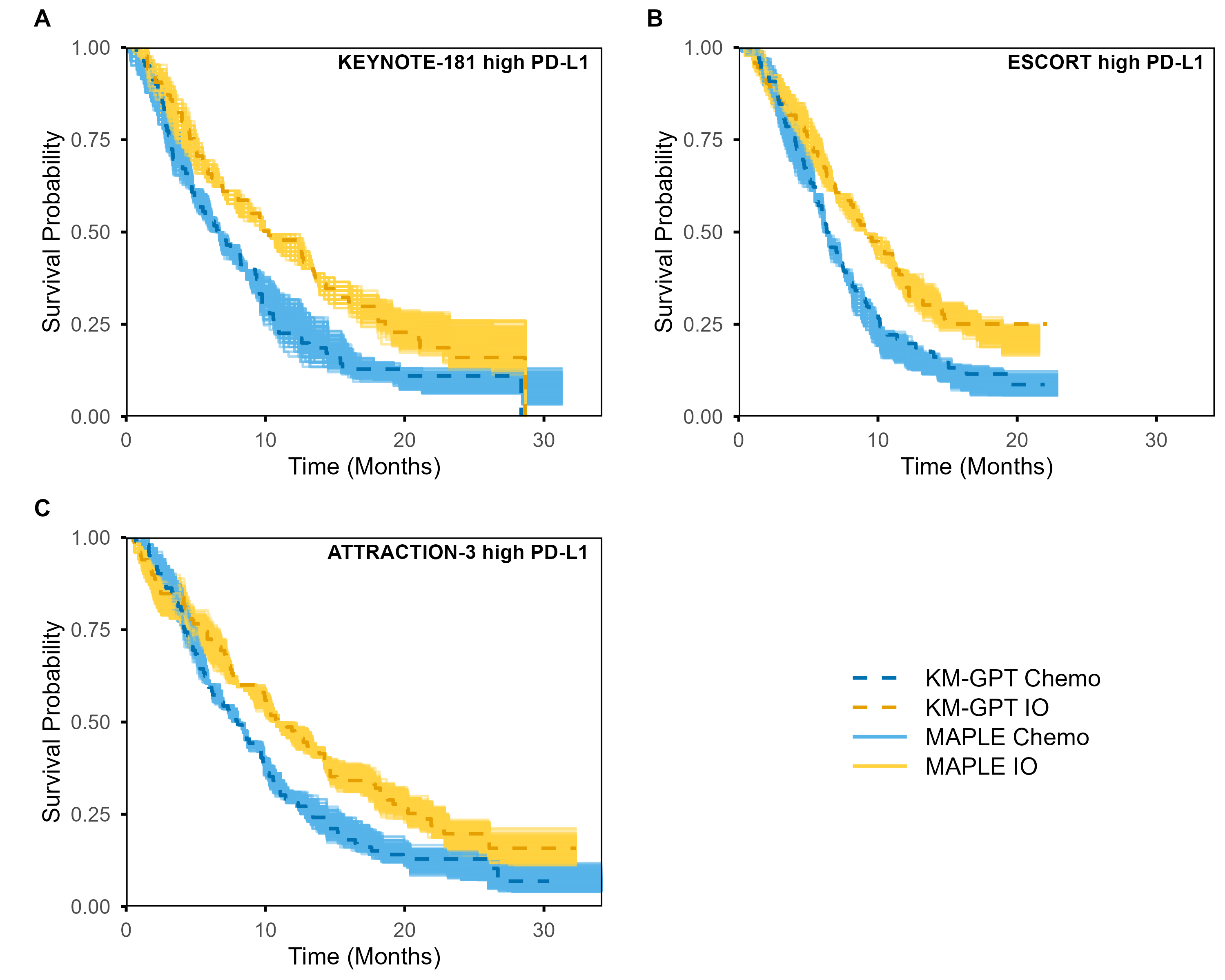}
    \caption{(A) KEYNOTE-181 PD-L1–high subgroups reconstructed by MAPLE. (B) ESCORT PD-L1–high subgroups reconstructed by MAPLE. (C) ATTRACTION-3 PD-L1–high subgroups reconstructed by MAPLE. IO = Immunotherapy, Chemo = Chemotherapy.}
    \label{fig_meta_overlay_km_high}
\end{figure}

\begin{table}[H]
\centering
\caption{Reconstruction error comparison, measured as root mean square error (RMSE) between the estimated survival probabilities and the simulated ground truth. RMSE is averaged across 100 evenly spaced time points from 0 to the maximum observation time. The table compares different IPD reconstruction methods across the Control and Treatment arms, including the overall population and subgroups stratified by high and low biomarker values.}
\renewcommand{\arraystretch}{1.2}
\setlength{\tabcolsep}{8pt}
\begin{tabular}{llcccccc}
\hline
 &  & \multicolumn{3}{c}{\textbf{Control}} & \multicolumn{3}{c}{\textbf{Treatment}} \\ \cline{1-8}
 \textbf{Biomarker Subgroup:} & & Overall & High & Low & Overall & High & Low \\ 
\hline
\multicolumn{8}{l}{\textbf{Reconstruction from KM Plots}} \\ 
\hline
{\bf VEC-KM + CEN-KM} &  & 0    & 0    & 0    & 0    & 0    & 0    \\
VEC-KM + IPDfromKM        &  & 0.005 & 0.007 & 0.009 & 0.004 & 0.006 & 0.008 \\
VEC-KM + KMtoIPD          &  & 0.002 & 0.003 & 0.003 & 0.003 & 0.005 & 0.005 \\ 
\hline
\multicolumn{8}{l}{\textbf{Subtraction}} \\ 
\hline
{\bf VEC-KM + CEN-KM} &  & - & - & 0 & - & - & 0 \\
VEC-KM + IPDfromKM        &  & - & - & 0.009 & - & - & 0.008 \\
VEC-KM + KMtoIPD          &  & - & - & 0.005 & - & - & 0.005 \\ 
\hline
\end{tabular}
\label{tab_sim_reconstruction_error}
\end{table}

\begin{table}[H]
\centering
\caption{Comparison of reported versus reconstructed median overall survival (mOS) values for immunotherapy and chemotherapy arms across four clinical trials. The mOS values are presented in months along with their corresponding 95\% confidence intervals (CI). }
\begin{tabular}{llc}
\hline
\textbf{Trial Name} & \textbf{Source} & \textbf{Overall (months (95\% CI))} \\
\hline
\multicolumn{3}{c}{\textbf{Immunotherapy Arms}} \\
\hline
RATIONALE-302 & Reported & 8.6 (7.5--10.4) \\
              & Reconstructed & 8.6 (7.5--10.4) \\
ATTRACTION-3  & Reported & 10.9 (9.2--13.3) \\
              & Reconstructed & 10.9 (9.2--13.3) \\
ESCORT        & Reported & 8.3 (6.8--9.7) \\
              & Reconstructed & 8.3 (6.8--9.7) \\
KEYNOTE-181   & Reported & 8.2 (6.7--10.3) \\
              & Reconstructed & 8.2 (6.7--10.3) \\
\hline
\multicolumn{3}{c}{\textbf{Chemotherapy Arms}} \\
\hline
RATIONALE-302 & Reported & 6.3 (5.3--7.0) \\
              & Reconstructed & 6.3 (5.3--7.0) \\
ATTRACTION-3  & Reported & 8.4 (7.2--9.9) \\
              & Reconstructed & 8.4 (7.2--9.9) \\
ESCORT        & Reported & 6.2 (5.7--6.9) \\
              & Reconstructed & 6.2 (5.7--6.9) \\
KEYNOTE-181   & Reported & 7.1 (6.1--8.2) \\
              & Reconstructed & 7.1 (6.1--8.2) \\
\hline
\end{tabular}
\label{tab_meta_overall_assess}
\end{table}

\begin{table}[H]
\centering
\caption{Summary of information extracted from original publications for MAPLE from KEYNOTE-181, ESCORT, and ATTRACTION-3 trials used in the subgroup meta-analysis. Each column represents summary statistics for chemotherapy (Chemo) and immunotherapy (IO) across studies. Total patient numbers, event counts, median overall survival (OS) with 95\% confidence interval (CI), and hazard ratios (HRs) comparing immunotherapy to chemotherapy within each PD-L1 subgroup (low, high, and unknown) are reported. Missing data, indicated by ``–", were not available in the respective publication.}
\renewcommand{\arraystretch}{1.25}
\setlength{\tabcolsep}{3pt}
\resizebox{\textwidth}{!}{
\begin{tabular}{@{}lcccccc@{}}
\toprule
 & \multicolumn{2}{c}{\textbf{KEYNOTE-181}} & \multicolumn{2}{c}{\textbf{ESCORT}} & \multicolumn{2}{c}{\textbf{ATTRACTION-3}} \\
\cmidrule(lr){2-3} \cmidrule(lr){4-5} \cmidrule(l){6-7}
 & \textbf{Chemo} & \textbf{IO} & \textbf{Chemo} & \textbf{IO} & \textbf{Chemo} & \textbf{IO} \\
\midrule
\multicolumn{7}{@{}l}{\textbf{PD-L1 Low}}\\
\quad Total patients & 119 & 109 & 118 & 129 & 107 & 109\\
\quad Events & \multicolumn{2}{c}{202} & 102 & 100 & 84 & 83\\
\quad Median OS (95\% CI) & 7.5 (6.3–8.3) & 7.3 (5.7–9.2) & \multicolumn{2}{c}{--} & 9.3 (7.2–12.0) & 10.9 (8.4–13.9)\\
\quad HR (95\% CI) & \multicolumn{2}{c}{0.88 (0.66–1.16)} & \multicolumn{2}{c}{0.82 (0.62–1.09)} & \multicolumn{2}{c}{0.84 (0.62–1.14)}\\[3pt]

\multicolumn{7}{@{}l}{\textbf{PD-L1 High}}\\
\quad Total patients & 82 & 85 & 98 & 93 & 102 & 101\\
\quad Events & \multicolumn{2}{c}{140} & 85 & 68 & 89 & 77\\
\quad Median OS (95\% CI) & 6.7 (4.8–8.6) & 10.3 (7.0–13.5) & 6.3 (5.5–7.5) & 9.2 (7.0–11.2) & 8.1 (6.0–9.9) & 10.9 (8.0–14.2)\\
\quad HR (95\% CI) & \multicolumn{2}{c}{0.64 (0.46–0.90)} & \multicolumn{2}{c}{0.58 (0.42–0.81)} & \multicolumn{2}{c}{0.69 (0.51–0.94)}\\[3pt]

\multicolumn{7}{@{}l}{\textbf{PD-L1 Unknown}}\\
\quad Total patients & 2 & 4 & 4 & 6 & \multicolumn{2}{c}{--}\\
\quad Events & \multicolumn{2}{c}{6} & 4 & 4 & \multicolumn{2}{c}{--}\\
\quad Median OS (95\% CI) & \multicolumn{2}{c}{--} & \multicolumn{2}{c}{--} & \multicolumn{2}{c}{--}\\
\quad HR (95\% CI) & \multicolumn{2}{c}{--} & \multicolumn{2}{c}{--} & \multicolumn{2}{c}{--}\\
\bottomrule
\end{tabular}
}
\\[3pt]
\label{tab_meta_maple_info}
\end{table}

\begin{table}[H]
\centering
\small
\caption{Summary of median overall survival (mOS) estimates and their corresponding confidence interval (CI) bounds calculated using 500 Monte Carlo simulations. Metrics reported for each treatment group (Chemotherapy and Immunotherapy) include the minimum, mean, and maximum values for mOS estimates, as well as the lower and upper bounds of the 95\% CI.}
\renewcommand{\arraystretch}{1.15}
\setlength{\tabcolsep}{6pt}
\begin{tabularx}{\textwidth}{c c *{3}{>{\centering\arraybackslash}X}}
\toprule
\textbf{Treatment} & \textbf{Metric} & \textbf{Min} & \textbf{Mean} & \textbf{Max} \\
\midrule
\multirow{3}{*}{Chemotherapy}  & Estimate & 7.21 & 7.33 & 7.47 \\
                               & Lower CI & 5.13 & 5.28 & 5.44 \\
                               & Upper CI & 8.97 & 9.21 & 9.45 \\
\cmidrule(lr){2-5}
\multirow{3}{*}{Immunotherapy} & Estimate & 8.02 & 8.25 & 8.44 \\
                               & Lower CI & 5.49 & 5.71 & 5.96 \\
                               & Upper CI & 10.42 & 10.74 & 11.05 \\
\bottomrule
\end{tabularx}
\label{tab_meta_median_mc_summary}
\end{table}

\begin{table}[H]
\centering
\small
\caption{Summary of hazard ratio (HR) estimates and corresponding confidence interval (CI) bounds derived from 500 Monte Carlo simulations. HR is defined as immunotherapy (IO) versus chemotherapy (Chemo). Results are reported across different time intervals (in months), including the minimum, mean, and maximum values for HR estimates, as well as the lower and upper bounds of the 95\% CI.}
\renewcommand{\arraystretch}{1.15}
\setlength{\tabcolsep}{6pt}
\begin{tabularx}{\textwidth}{c c *{3}{>{\centering\arraybackslash}X}}
\toprule
\textbf{Time Interval (months)} & \textbf{Metric} & \textbf{Min} & \textbf{Mean} & \textbf{Max} \\
\midrule
\multirow{3}{*}{0--6}   & Estimate & 0.87 & 0.93 & 0.99 \\
                        & Lower CI & 0.79 & 0.84 & 0.90 \\
                        & Upper CI & 0.95 & 1.01 & 1.10 \\
\cmidrule(lr){2-5}
\multirow{3}{*}{6--12}  & Estimate & 0.66 & 0.75 & 0.84 \\
                        & Lower CI & 0.58 & 0.66 & 0.76 \\
                        & Upper CI & 0.74 & 0.84 & 0.93 \\
\cmidrule(lr){2-5}
\multirow{3}{*}{12--18} & Estimate & 0.70 & 0.86 & 1.07 \\
                        & Lower CI & 0.59 & 0.74 & 0.91 \\
                        & Upper CI & 0.82 & 1.00 & 1.28 \\
\cmidrule(lr){2-5}
\multirow{3}{*}{18--24} & Estimate & 0.76 & 1.15 & 2.24 \\
                        & Lower CI & 0.51 & 0.86 & 1.57 \\
                        & Upper CI & 0.93 & 1.61 & 3.92 \\
\cmidrule(lr){2-5}
\multirow{3}{*}{24--30} & Estimate & 0.40 & 0.92 & 1.97 \\
                        & Lower CI & 0.17 & 0.62 & 1.30 \\
                        & Upper CI & 0.58 & 1.30 & 4.45 \\
\cmidrule(lr){2-5}
\multirow{3}{*}{30--36} & Estimate & 0.69 & 1.08 & 1.81 \\
                        & Lower CI & 0.40 & 0.72 & 1.28 \\
                        & Upper CI & 0.90 & 1.57 & 3.09 \\
\bottomrule
\end{tabularx}
\label{tab_meta_hr_mc_summary}
\end{table}


\begin{thebibliography}{10}

\bibitem{debray2015individual}
Thomas~PA Debray, Richard~D Riley, Maroeska~M Rovers, Johannes~B Reitsma, Karel~GM Moons, and Cochrane IPD~Meta analysis Methods~group.
\newblock Individual participant data (ipd) meta-analyses of diagnostic and prognostic modeling studies: guidance on their use.
\newblock {\em PLoS medicine}, 12(10):e1001886, 2015.

\bibitem{smith2016individual}
Catrin~Tudur Smith, Maura Marcucci, Sarah~J Nolan, Alfonso Iorio, Maria Sudell, Richard Riley, Maroeska~M Rovers, and Paula~R Williamson.
\newblock Individual participant data meta-analyses compared with meta-analyses based on aggregate data.
\newblock {\em Cochrane Database of Systematic Reviews}, (9), 2016.

\bibitem{riley2010meta}
Richard~D Riley, Paul~C Lambert, and Ghada Abo-Zaid.
\newblock Meta-analysis of individual participant data: rationale, conduct, and reporting.
\newblock {\em Bmj}, 340, 2010.

\bibitem{batson2016review}
Sarah Batson, Gemma Greenall, and Pollyanna Hudson.
\newblock Review of the reporting of survival analyses within randomised controlled trials and the implications for meta-analysis.
\newblock {\em PLoS One}, 11(5):e0154870, 2016.

\bibitem{guyot2012}
Patricia Guyot, A.~E. Ades, Mario J. N.~M. Ouwens, and Nicky~J. Welton.
\newblock Enhanced secondary analysis of survival data: reconstructing the data from published {K}aplan--{M}eier survival curves.
\newblock {\em BMC Medical Research Methodology}, 12:9, 2012.

\bibitem{Liu2021}
Na~Liu, Yanhong Zhou, and J.~Jack Lee.
\newblock {IPDfromKM}: reconstruct individual patient data from published kaplan–meier survival curves.
\newblock {\em BMC Medical Research Methodology}, 21(1):111, 2021.

\bibitem{zhao2022kmsubtraction}
Joseph~J Zhao, Nicholas~L Syn, Benjamin Kye~Jyn Tan, Dominic Wei~Ting Yap, Chong~Boon Teo, Yiong~Huak Chan, and Raghav Sundar.
\newblock Kmsubtraction: reconstruction of unreported subgroup survival data utilizing published kaplan-meier survival curves.
\newblock {\em BMC medical research methodology}, 22(1):93, 2022.

\bibitem{rogula2022method}
Basia Rogula, Greta Lozano-Ortega, and Karissa~M Johnston.
\newblock A method for reconstructing individual patient data from kaplan-meier survival curves that incorporate marked censoring times.
\newblock {\em MDM Policy \& Practice}, 7(1):23814683221077643, 2022.

\bibitem{zhao2025synthipd}
Zixuan Zhao, Zexin Ren, Guannan Zhai, Feifang Hu, Will Ma, En~Xie, and Qian Shi.
\newblock Synthipd: assumption-lean synthetic individual patient data generation.
\newblock {\em arXiv preprint arXiv:2509.16466}, 2025.

\bibitem{ZhaoSunDingXu2025}
Yao Zhao, Haoyue Sun, Yantian Ding, and Yanxun Xu.
\newblock Km-gpt: An automated pipeline for reconstructing individual patient data from kaplan-meier plots, 2025.
\newblock \url{https://arxiv.org/abs/2509.18141}.

\bibitem{yu2022association}
Yao Yu, Kaveh Zakeri, Eric~J Sherman, and Nancy~Y Lee.
\newblock Association of low and intermediate combined positive scores with outcomes of treatment with pembrolizumab in patients with recurrent and metastatic head and neck squamous cell carcinoma: secondary analysis of keynote 048.
\newblock {\em JAMA oncology}, 8(8):1216--1218, 2022.

\bibitem{burtness2019pembrolizumab}
Barbara Burtness, Kevin~J Harrington, Richard Greil, Denis Souli{\`e}res, Makoto Tahara, Gilberto de~Castro, Amanda Psyrri, Neus Bast{\'e}, Prakash Neupane, {\AA}se Bratland, et~al.
\newblock Pembrolizumab alone or with chemotherapy versus cetuximab with chemotherapy for recurrent or metastatic squamous cell carcinoma of the head and neck (keynote-048): a randomised, open-label, phase 3 study.
\newblock {\em The Lancet}, 394(10212):1915--1928, 2019.

\bibitem{schoenfeld2020keynote}
Jonathan~D Schoenfeld, Geoffrey Fell, Robert~I Haddad, and Lorenzo Trippa.
\newblock Keynote 48: Is it really for everyone?
\newblock {\em medRxiv}, pages 2020--04, 2020.

\bibitem{shen2022tislelizumab}
Lin Shen, Ken Kato, Sung-Bae Kim, Jaffer~A Ajani, Kuaile Zhao, Zhiyong He, Xinmin Yu, Yongqian Shu, Qi~Luo, Jufeng Wang, et~al.
\newblock Tislelizumab versus chemotherapy as second-line treatment for advanced or metastatic esophageal squamous cell carcinoma (rationale-302): a randomized phase iii study.
\newblock {\em Journal of clinical oncology}, 40(26):3065--3076, 2022.

\bibitem{doi2016keynote}
Toshihiko Doi, Jaafar Bennouna, Lin Shen, Peter~C Enzinger, Ruixue Wang, Ildiko Csiki, Minori Koshiji, and Manish~A Shah.
\newblock Keynote-181: Phase 3, open-label study of second-line pembrolizumab vs single-agent chemotherapy in patients with advanced/metastatic esophageal adenocarcinoma., 2016.

\bibitem{huang2020camrelizumab}
Jing Huang, Jianming Xu, Yun Chen, Wu~Zhuang, Yiping Zhang, Zhendong Chen, Jia Chen, Helong Zhang, Zuoxing Niu, Qingxia Fan, et~al.
\newblock Camrelizumab versus investigator's choice of chemotherapy as second-line therapy for advanced or metastatic oesophageal squamous cell carcinoma (escort): a multicentre, randomised, open-label, phase 3 study.
\newblock {\em The Lancet Oncology}, 21(6):832--842, 2020.

\bibitem{kato2019nivolumab}
Ken Kato, Byoung~Chul Cho, Masanobu Takahashi, Morihito Okada, Chen-Yuan Lin, Keisho Chin, Shigenori Kadowaki, Myung-Ju Ahn, Yasuo Hamamoto, Yuichiro Doki, et~al.
\newblock Nivolumab versus chemotherapy in patients with advanced oesophageal squamous cell carcinoma refractory or intolerant to previous chemotherapy (attraction-3): a multicentre, randomised, open-label, phase 3 trial.
\newblock {\em The Lancet Oncology}, 20(11):1506--1517, 2019.

\bibitem{yap2023}
Dominic W.~T. Yap, Alberto~G. Leone, Nicky Z.~H. Wong, Joseph~J. Zhao, Jeremy C.~S. Tey, Raghav Sundar, and Filippo Pietrantonio.
\newblock Effectiveness of immune checkpoint inhibitors in patients with advanced esophageal squamous cell carcinoma: A meta-analysis including low pd-l1 subgroups.
\newblock {\em JAMA Oncology}, 9(2):215--224, 2023.

\end{thebibliography}

\begin{thebibliography}{1}

\bibitem{DigitizeIt}
I.~Bormann.
\newblock Digitizeit.

\bibitem{WebPlotDigitizer}
Ankit Rohatgi.
\newblock Webplotdigitizer.

\bibitem{Zhang2024}
Jasper~Z. Zhang, Juan~D. Rios, Thanos Pechlivanoglou, Willem Witteman, Jaclyn Beca, Jeffrey Hoch, and Murray~D. Krahn.
\newblock Survdigitizer: an algorithm for automated survival curve digitization.
\newblock {\em BMC Medical Research Methodology}, 24:147, 2024.

\bibitem{ZhaoSunDingXu2025_supp}
Yao Zhao, Haoyue Sun, Yantian Ding, and Yanxun Xu.
\newblock Km-gpt: An automated pipeline for reconstructing individual patient data from kaplan-meier plots, 2025.
\newblock \url{https://arxiv.org/abs/2509.18141}.

\bibitem{PyMuPDF}
Artifex~Software Inc.
\newblock Pymupdf — python binding for mupdf, 2025.
\newblock Accessed: 2025-10-28.

\bibitem{zhao2022kmsubtraction_supp}
Joseph~J Zhao, Nicholas~L Syn, Benjamin Kye~Jyn Tan, Dominic Wei~Ting Yap, Chong~Boon Teo, Yiong~Huak Chan, and Raghav Sundar.
\newblock Kmsubtraction: reconstruction of unreported subgroup survival data utilizing published kaplan-meier survival curves.
\newblock {\em BMC medical research methodology}, 22(1):93, 2022.

\bibitem{kato2019nivolumab_supp}
Ken Kato, Byoung~Chul Cho, Masanobu Takahashi, Morihito Okada, Chen-Yuan Lin, Keisho Chin, Shigenori Kadowaki, Myung-Ju Ahn, Yasuo Hamamoto, Yuichiro Doki, et~al.
\newblock Nivolumab versus chemotherapy in patients with advanced oesophageal squamous cell carcinoma refractory or intolerant to previous chemotherapy (attraction-3): a multicentre, randomised, open-label, phase 3 trial.
\newblock {\em The Lancet Oncology}, 20(11):1506--1517, 2019.

\bibitem{parejo2012metaheuristic}
Jos{\'e}~Antonio Parejo, Antonio Ruiz-Cort{\'e}s, Sebasti{\'a}n Lozano, and Pablo Fernandez.
\newblock Metaheuristic optimization frameworks: a survey and benchmarking.
\newblock {\em Soft Computing}, 16(3):527--561, 2012.

\bibitem{kirkpatrick1983optimization}
Scott Kirkpatrick, C~Daniel Gelatt~Jr, and Mario~P Vecchi.
\newblock Optimization by simulated annealing.
\newblock {\em science}, 220(4598):671--680, 1983.

\end{thebibliography}
\end{document}